\RequirePackage{fix-cm}
\documentclass[natbib,smallextended]{svjour3}       % onecolumn (second format)
\smartqed  % flush right qed marks, e.g. at end of proof

\usepackage{amsmath,amssymb,amsfonts}
\usepackage{algorithmic}
\usepackage{graphicx}
\usepackage{textcomp}

\usepackage{txfonts}
\usepackage{mathdots}

\usepackage{algorithm}
\usepackage{algorithmic}

  \usepackage{rotating}
        \usepackage{multirow}
        \usepackage{lscape}

\usepackage{xcolor}

 \journalname{}
\begin{document}

\title{ Latent Dirichlet Allocation (LDA) and Topic modeling: models, applications, a survey
}


\author{Hamed Jelodar \and Yongli Wang \and Chi Yuan \and Xia Feng  \and Xiahui Jiang \and Yanchao Li \and Liang Zhao %etc.
}


\institute{H. Jelodar \at
              School of Computer Science and Technology, Nanjing University of Science and Technology, Nanjing 210094, China \\
              Tel.: +8615298394479\\
              \email{Jelodar@njust.edu.cn}           %  \\
           \and
           Y. Wang \at
              School of Computer Science and Technology, Nanjing University of Science and Technology, Nanjing 210094, China \\
              \email{YongliWang@njust.edu.cn}           %  \\
}

\date{Received: date / Accepted: date}

\maketitle

\begin{abstract}

Topic modeling is one of the most powerful techniques in text mining for data mining, latent data discovery, and finding relationships among data and text documents. Researchers have published many articles in the field of topic modeling and applied in various fields such as software engineering, political science, medical and linguistic science, etc. There are various methods for topic modelling; Latent Dirichlet Allocation (LDA) is one of the most popular in this field. Researchers have proposed various models based on the LDA in topic modeling. According to previous work, this paper will be very useful and valuable for introducing LDA approaches in topic modeling. In this paper, we investigated highly scholarly articles (between 2003 to 2016) related to topic modeling based on LDA to discover the research development, current trends and intellectual structure of topic modeling. In addition, we summarize challenges and introduce famous tools and datasets in topic modeling based on LDA.

 \keywords{ Topic modeling, Latent Dirichlet Allocation, tag recommendation, Semantic web, Gibbs Sampling }
\end{abstract}

\section{Introduction}
\label{sec1}

Natural language processing (NLP) is a challenging research in computer science to information management, semantic mining, and enabling computers to obtain meaning from human language processing in text-documents. Topic modeling methods are powerful smart techniques that widely applied in natural language processing to topic discovery and semantic mining from unordered documents [1]. In a wide perspective, Topic modeling methods based on LDA  have been applied to natural language processing, text mining, and social media analysis, information retrieval. For example, topic modeling based on social media analytics facilitates understanding the reactions and conversations between people in online communities, As well as extracting useful patterns and understandable from their interactions in addition to what they share on social media websites such as twiiter facebook [2-3]. Topic models are prominent for demonstrating discrete data; also, give a productive approach to find hidden structures(semantics) in gigantic information. There are many papers for in this field and definitely cannot mention to all of them, so we selected more signification papers. Topic models are applied in various fields including medical sciences  [4-7]  , software engineering [8-12], geography [13-17], political science [18-20] , etc.\\

For example in political science, In [20] proposed a new two-layer matrix factorization methodology for identifying topics in large political speech corpora over time and identify both niche topics related to events at a particular point in time and broad, long-running topics. This paper has focused on European Parliament speeches, the proposed topic modeling method has a number of potential applications in the study of politics, including the analysis of speeches in other parliaments, political manifestos, and other more traditional forms of political texts. In [21] suggested a new unsupervised topic model based on LDA for contrastive opinion modeling which purpose to find the opinions from multiple views, according to a given topic and their difference on the topic with qualifying criteria, the model called Cross-Perspective Topic (CPT) model. They performed experiments with both qualitative and quantitative measures on two datasets in the political area that include: first dataset is statement records of U.S. senators that show political stances of senators. Also for the second dataset, extracted of world News Medias from three representative media in U.S (New York Times), China (Xinhua News) and India (Hindu). To evaluate their approach with other models, used corrIDA and LDA as two baselines.\\

Another group of researchers focused on topic modeling in software engineering, in [8] for the first time, they used LDA, to extract topics in source code and perform visualization of software similarity, In other words, LDA is used as an intuitive approach for calculation of similarity between source files and obtain their respective distributions of each document over topics. They utilized their method on 1,555 software projects from Apache and SourceForge that includes 19 million source lines of code (SLOC). The authors demonstrated this approach, can be effective for project organization, software refactoring. In [22] introduced a method based on LDA for automatically categorizing software systems, called LACT. For evaluation of LACT, used 43 open-source software systems in different programming languages and showed LACT can categorize software systems based on type of programming language. In [23, 24] proposed an approach topic modeling based on LDA model for the purpose of bug localization. Their idea, applied to analysis of same bugs in Mozilla and Eclipse and result showed that their LDA-based approach is better than LSI for evaluate and analyze bugs in these source codes.\\

An analysis of geographic information is another issue that can be referred to [17].  They introduced a novel method based on multi-modal Bayesian models to describe social media by merging text features and spatial knowledge that called GeoFolk. As a general outlook, this method can be considered as an extension of Latent Dirichlet Allocation (LDA). They used the available standard CoPhIR dataset that contains an abundance of over 54 million Flickr. The GeoFolk model has the ability to be used in quality-oriented applications and can be merged with some models from Web 2.0 social. In [16], this article examines the issue of topic modeling to extract the topics from geographic information and GPS-related documents. They suggested a new location text method that is a combination of topic modeling and geographical clustering called LGTA (Latent Geographical Topic Analysis). To test their approaches, they collected a set of data from the website Flickr, according to various topics.\\

In other view, Most of the papers that were studied, had goals for this topic modeling, such as: Source code analysis
[8 , 9, 22, 24-27] , Opinion and aspect Mining [18, 28-36], Event detection [37-40], Image classification [13, 41], recommendation system [42-48] and emotion classification [49-51], etc. For example in recommendation system, in [44] proposed a personalized hashtag recommendation approach based LDA model that can discover latent topics in microblogs, called Hashtag-LDA and applied experiments on "UDI-TwitterCrawl-Aug2012-Tweets" as a real-world Twitter dataset.

\subsection{ literature review and related works}
Topic models have many applications in natural processing languages. Many articles have been published based on topic modeling approaches in various subject such as Social Network, software engineering, Linguistic science and etc. There are some works that have focused on survey in Topic modeling.\\
In [52], the authors presented a survey on topic modeling in software engineering field to specify how topic models have thus far been applied to one or more software repositories. They focused on articles written between Dec 1999 to Dec 2014 and surveyed 167 article that using topic modeling in software engineering area. They identified and demonstrate the research trends in mining unstructured repositories by topic models. They found that most of studies focused on only a limited number of software engineering task and also most studies use only basic topic models. In [53], the authors focused on survey in Topic Models with soft clustering abilities in text corpora and investigated basic concepts and existing models classification in various categories with parameter estimation (such as Gibbs Sampling) and performance evaluation measures. In addition, the authors presented some applications of topic models for modeling text corpora and discussed several open issues and future directions.\\

In [54], introduced and surveyed the field of opinion mining and sentiment analysis, which helps us to observe a elements from the intimidating unstructured text. The authors discussed the most extensively studied subject of subjectivity classification and sentiment which specifies whether a document is opinionated. Also, they described aspect-based sentiment analysis which exploits the full power of the abstract model and discussed about aspect extraction based on topic modeling approaches. In [55], they discussed challenges of text mining techniques in information systems research and indigested the practical application of topic modeling in combination with explanatory regression analysis, using online customer reviews as an exemplary data source.\\

In [56], the authors presented a survey of how topic models have thus far been applied in Software Engineering tasks from four SE journals and eleven conference proceedings. They considered 38 selected publications from 2003 to 2015 and found that topic models are widely used in various SE tasks in an increasing tendency, such as social software engineering, developer recommendation and etc. Our research difference with other works is that, we had a deep study on topic modeling approaches based on LDA with the coverage of various aspects such  as applications, tools , dataset and models.

\subsection{Motivations and contributions}
Since, topic modeling techniques can be very useful and effective in natural language processing to semantic mining and latent discovery in documents and datasets. Hence, our motivation is to investigate the topic modeling approaches in different subjects with the coverage of various aspects such as models, tools, dataset and applications. The main goal of this work is to provide an overview of the methods of topic modeling based on LDA. In summary, this paper makes four main contributions:

\begin{itemize}
\item We investigate scholarly articles (from 2003 to 2016) which are related to Topic Modeling based on LDA to discover the research development, current trends and intellectual structure of topic modeling based on LDA.
\item We investigate topic modeling applications in various sciences.
\item We summarize challenges in topic modeling, such as image processing, Visualizing topic models, Group discovery, User Behavior Modeling, and etc.
\item We introduce some of the most famous data and tools in topic modeling.
\end{itemize}

\section{Computer science and topic modeling}

Topic models have an important role in computer science for text mining and natural language processing. In Topic modeling, a topic is a list of words that occur in statistically significant methods. A text can be an email, a book chapter, a blog posts, a journal article and any kind of unstructured text. Topic models cannot understand the means and concepts of words in text documents for topic modeling. Instead, they suppose that any part of the text is combined by selecting words from probable baskets of words where each basket corresponds to a topic. The tool goes via this process over and over again until it stays on the most probable distribution of words into baskets which call topics. Topic modeling can provide a useful view of a large collection in terms of the collection as a whole, the individual documents, and relationships between documents. In figure 1, we provided a taxonomy of topic modeling methods based on LDA, from some of the impressive works.

\subsection{Latent Dirichlet Allocation}
LDA is a generative probabilistic model of a corpus. The basic idea is that the documents are represented as random mixtures over latent topics, where a topic is characterized by a distribution over words. Latent Dirichlet allocation (LDA), first introduced by Blei, Ng and Jordan in 2003 [1], is one of the most popular methods in topic modeling. LDA represents topics by word probabilities. The words with highest probabilities in each topic usually give a good idea of what the topic is can word probabilities from LDA.\\

LDA, an unsupervised generative probabilistic method for modeling a corpus, is the most commonly used topic modeling method. LDA assumes that each document can be represented as a probabilistic distribution over latent topics, and that topic distribution in all documents share a common Dirichlet prior. Each latent topic in the LDA model is also represented as a probabilistic distribution over words and the word distributions of topics share a common Dirichlet prior as well. Given a corpus $D$ consisting of $M$ documents, with document $d$ having ~\textit{N}${}_{d}$ words ($d$ $\mathrm{\in }$ {1,..., $M$}), LDA models $D$ according to the following generative process [4]:

\noindent \textit{ (a) } Choose a multinomial distribution~\textit{$\varphi$}${}_{~}$\textit{${}_{t}$}${}_{~}$for topic~\textit{t}~(\textit{t}~$\mathrm{\in }$$\{$1,...,~\textit{T}$\}$) from a Dirichlet distribution with parameter~\textit{$\beta$}

\noindent \textit{ (b) } Choose a multinomial distribution~\textit{$\theta$}${}_{~}$\textit{${}_{d}$}${}_{~}$for document~\textit{d}~(\textit{d}~$\mathrm{\in }$$\{$1,...,~\textit{M}$\}$) from a Dirichlet distribution with parameter~\textit{$\alpha$}.

\noindent \textit{ (c) } For a word~\textit{w}${}_{~}$\textit{${}_{n}$}${}_{~}$(\textit{n}~$\mathrm{\in }$$\{$1,...,~\textit{N}${}_{~}$\textit{${}_{d}$}${}_{~}$$\}$) in document~\textit{d},

\begin{enumerate}
\item \begin{enumerate}
\item \begin{enumerate}
\item  Select a topic~\textit{z}${}_{~}$\textit{${}_{n}$}${}_{~}$from~\textit{$\theta$}${}_{~}$\textit{${}_{d}$}${}_{~}$.

\item  \textit{  }Select a word~\textit{w}${}_{~}$\textit{${}_{n}$}${}_{~}$from~\textit{$\varphi$}${}_{~}$\textit{${}_{zn}$}${}_{~}$.
\end{enumerate}
\end{enumerate}
\end{enumerate}

\noindent In above generative process, words in documents are only observed variables while others are latent variables (\textit{$\varphi$}~and~\textit{$\theta$}) and hyper parameters (\textit{$\alpha$}~and~\textit{$\beta$}). In order to infer the latent variables and hyper parameters, the probability of observed data~\textit{D}~is computed and maximized as follows:

\begin{equation*} p(D\vert \alpha,\beta)=\prod_{d=1}^{M}\int p(\theta_{d}\vert \alpha)(\prod_{n=1}^{N_{d}} \sum_{z_{dn}}p(Z_{dn}\vert\theta_{d})p(w_{dn}\vert z_{dn},\beta))d\theta_{d}\tag{1} \end{equation*}

\noindent Defined $\alphaup$~parameters of topic Dirichlet prior and the distribution of words over topics, which, drawn from the Dirichlet distribution, given $\betaup$. Defined, $T$ is the number of topics, \textit{M} as number of documents; \textit{N} is the size of the vocabulary. The Dirichlet-multinomial pair for the corpus-level topic distributions, considered as (\textit{$\alpha$, $\theta$}).\textit{ }The Dirichlet-multinomial pair for topic-word distributions, given (\textit{$\beta$, }$\varphi $). The variables \textit{$\theta$}${}_{~}$\textit{${}_{d}$} are document-level variables, sampled when per document.  $z_{dn}\ ,w_{d_n}\ $ variables are word-level variables and are sampled when for each word in each text-document.

\noindent

     LDA is a distinguished tool for latent topic distribution for a large corpus. Therefore, it has the ability to identify sub-topics for a technology area composed of many patents, and represent each of the patents in an array of topic distributions. With LDA, the terms in the set of documents, generate a vocabulary that is then applied to discover hidden topics. Documents are treated as a mixture of topics, where a topic is a probability distribution over this set of terms. Each document is then seen as a probability distribution over set of topics. We can think of the data as coming from a generative process that is defined by the joint probability distribution over what is observed and what is hidden.

\subsection{Parameter estimation, Inference, Training for LDA}
Various methods have been proposed to estimate LDA parameters, such as variational method [1], expectation propagation [57] and Gibbs sampling [58].

\begin{itemize}
\item \textbf{Gibbs sampling}, is a Monte Carlo Markov-chain algorithm, powerful technique in statistical inference, and a method of generating a sample from a joint distribution when only conditional distributions of each variable can be efficiently computed. According to our knowledge, researchers have widely used this method for the LDA. Some of works related based on LDA and Gibbs, such as [22, 50, 59-66].
\item \textbf{Expectation-Maximization (EM)}, is a powerful method to obtain parameter estimation of graphical models and can use for unsupervised learning. In fact, the algorithm relies on discovering the maximum likelihood estimates of parameters when the data model depends on certain latent variables EM algorithm contains two steps, the E-step (expectation) and the M-step (maximization). Some researchers have applied this model to LDA training, such as  [67-70].
\item \textbf{Variational Bayes inference (VB)}, VB can be considered as a type of   EM extension that uses a parametric approximation to the posterior distribution of both parameters and other latent variables and attempts to optimize the fit (e.g. using KL-divergence) to the observed data. Some researchers have applied this model to LDA training, such as [71-72].
\end{itemize}

\begin{figure*}
 \includegraphics[scale=.7]{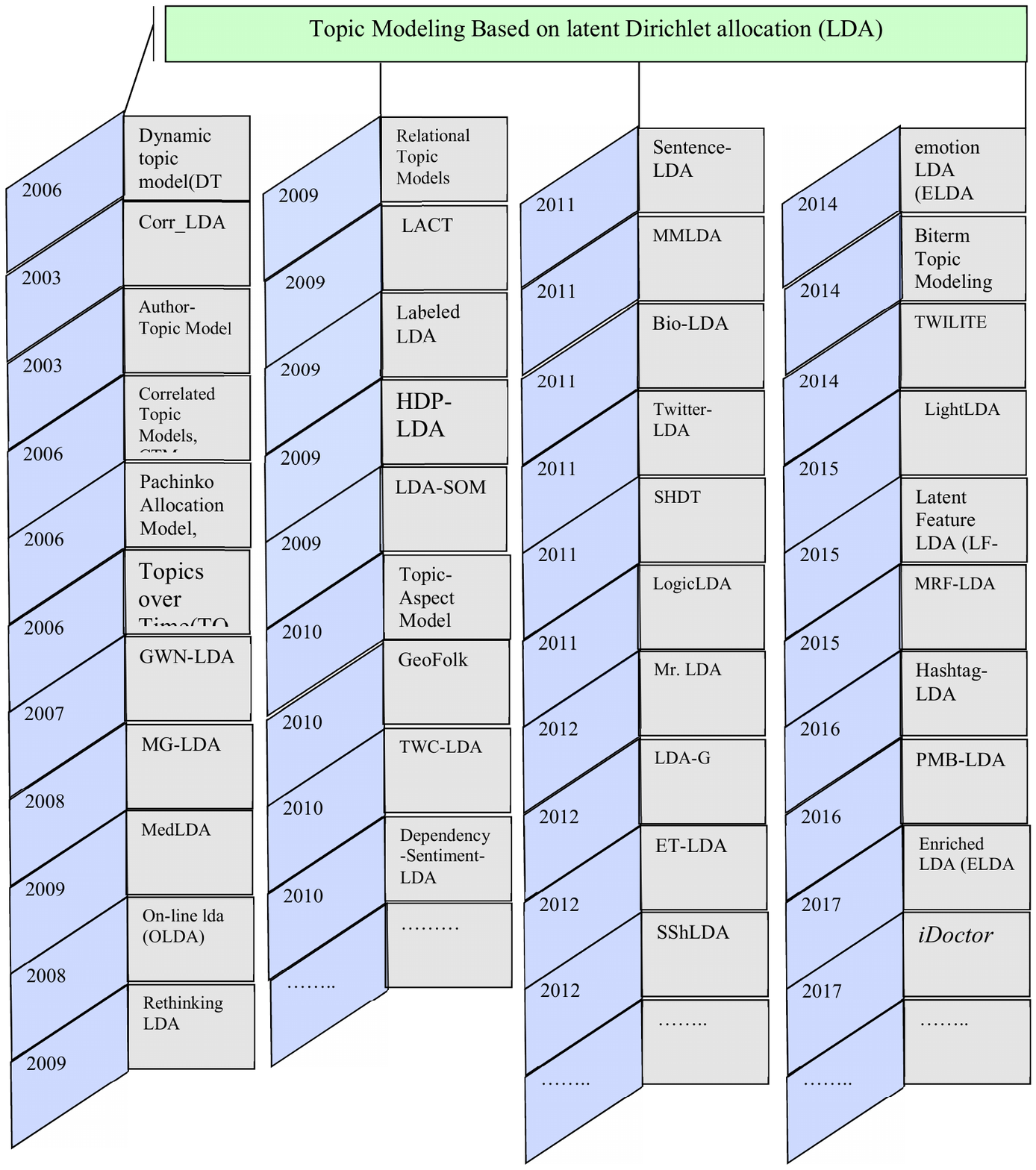}
\caption{A taxonomy of methods based on extension LDA, considered some of the impressive works}
\label{fig:1}     % Give a unique label
\end{figure*}

\subsection{ A brief look at past work: Research between 2003 to 2009}

The LDA was first presented in 2003, and researchers have been tried to provide extended approaches based on LDA, as shown in Table 1. Undeniably, this period (2003 to 2009) is very important because key and baseline approaches were introduced, such as: Corr\_LDA, Author-Topic Model , DTM and , RTM etc .\\

Author-Topic model [75], is a popular and simple probabilistic model in topic modeling for finding relationships among authors, topics, words and documents.  This model provides a distribution of different topics for each author and also a distribution of words for each topic. For evaluation, the authors used 1700 papers of NIPS conference and also 160,000 CiteSeer abstracts of CiteSeer dataset. To estimate the topic and author distributions applied Gibbs sampling. According to their result, showed this approach can provide a significantly predictive for interests of authors in perplexity measure. This model has attracted much attention from researchers and many approaches proposed based on ATM, such as [12, 76].\\

DTM, Dynamic Topic Model (DTM) is introduced by Blei and Lafferty as an extension of LDA that this model can obtain evolution of topics over time in a sequentially arranged corpus of documents and exhibits evolution of word-topic distribution which causes it easy to vision the topic trend [73]. As an advantage, DTM is very impressible for extracting topics from collections that change slowly over a period of time.\\
Labeled LDA (LLDA) is another LDA extension which suppose that each document has a set of known labels [66]. This model can be trained with labeled documents and even supports documents with more than one label. Topics are learned from the co-occurring terms in places from the same category, with topics approximately capturing different place categories. A separate L-LDA model is trained for each place category, and can be used to infer the category of new, previously unseen places. LLDA is a supervised algorithm that makes topics applying the Labels assigned manually. Therefore, LLDA can obtain meaningful topics, with words that map well to the labels applied. As a disadvantage, Labeled LDA has limitation to support latent subtopics within a determined label or any global topics. For overcome this problem, proposed partially labeled LDA (PLDA) [74].\\

MedLDA, proposed the maximum entropy discrimination latent Dirichlet allocation (MedLDA) model, which incorporates the mechanism behind the hierarchical Bayesian models (such as, LDA) with the max margin learning (such as SVMs) according to a unified restricted optimization framework. In fact each data sample is considered to a point in a finite dimensional latent space, of which each feature corresponds to a topic, i.e., unigram distribution over terms in a vocabulary [67]. MEDLDA model can be applied in both classification and regression. However, the quality of the topical space learned using MedLDA is heavily affected by the quality of the classifications which are learned at per iteration of MedLDA, as a disadvantage.\\

Relational Topic Models (RTM), is another extension, RTM is a hierarchical model of networks and per-node attribute data. First, each document was created from topics in LDA. Then, modelling the connections between documents and considered as binary variables, one for each pair from documents. These are distributed based on a distribution that depends on topics used to generate each of the constituent documents. So in this way, the content of the documents are statistically linked to the link structure between them and we can say that this model can be used to summarize a network of documents [69]. In fact, the advantage of RTM is that it considers both links and document context between documents. The disadvantage of RTM is its scalability. Since RTM can only make predictions for a document couple, this means that the action of recommending related questions for a recently created one would bring in computing couple RTM responses among every available question in the query and the collection.

\subsection{ A brief look at past work: Research between 2010 to 2011}
Twenty seven approaches are introduced in Table 1, where eight were published in 2010 and six in 2011. According to the Table 1, used LDA model for variety subjects, such as: Scientific topic discovery [35], Source code analysis [27], Opinion Mining [30], Event detection [40], Image Classification [41].\\

Sizov et al. in [17] introduced a novel method based on multi-modal Bayesian models to describe social media by merging text features and spatial knowledge that called GeoFolk. As a general outlook, this method can be considered as an extension of Latent Dirichlet Allocation (LDA). They used the available standard CoPhIR dataset that it contains an abundance of over 54 million Flickr. The GeoFolk model has ability to be used in quality-oriented applications and can be merged with some models from Web 2.0 social.\\

Z. Zhai et al. in [30] used prior knowledge as a constraint in the LDA models to improve grouping of features by LDA. They extract must link and cannot-link constraint from the corpus. Must link indicates that two features must be in the same group while cannot-link restricts that two features cannot be in the same group. These constraints are extracted automatically. If at least one of the terms of two product features are same, they are considered to be in the same group as must link. On the other hand, if two features are expressed in the same sentence without conjunction "and", they are considered as a different feature and should be in different groups as cannot-link.\\

 Wang et al. in [85] suggested an approach based on LDA that called Bio-LDA that can identify biological terminology to obtain latent topics. The authors have shown that this approach can be applied in different studies such as association search, association predication, and connectivity map generation. And they showed that Bio-LDA can be applied to increase the application of molecular bonding techniques as heat maps.\\

\begin{landscape}
\begin{table}[]
\centering
\caption{Some impressive articles based on LDA: between 2003 - 2011}
\label{my-label}

\resizebox{19cm}{!} {
\begin{tabular}{|l|l|l|l|l|l|}
\hline
\multicolumn{1}{|c|}{\textbf{Author-study}} & \multicolumn{1}{c|}{\textbf{Model}}           & \multicolumn{1}{c|}{\textbf{Years}} & \multicolumn{1}{c|}{\textbf{Parameter Estimation / Inference}} & \multicolumn{1}{c|}{\textbf{Methods}}       & \multicolumn{1}{c|}{\textbf{Problem Domain}}                \\ \hline

[70]             & Corr\_LDA                           & 2003           & Variational EM                                   & LDA     & Image annotation and retrieval          \\ \hline

[75]              & Author-Topic Model                  & 2004           & Gibbs Sampling                                   & LDA     & \begin{tabular}[c]{@{}l@{}}Find the relationships between authors,\\ documents,words, and topics\end{tabular} \\ \hline

[76]          & AuthorRecipient-Topic (ART)         & 2005           & Gibbs Sampling                                   & \begin{tabular}[c]{@{}l@{}}LDA\\ Author-Topic (AT)\end{tabular}               & Social network analysis and role discovery                                 \\ \hline

[73]             & Dynamic topic model(DTM)            & 2006           & Kalman variational algorithm                     & \begin{tabular}[c]{@{}l@{}}LDA\\ Galton Watson process\end{tabular}           & Provide a dynamic model for evolution of topics                            \\ \hline

[77]              & Topics over Time(TOT)               & 2006           & Gibbs Sampling                                   & LDA     & Capture word co-occurrences and localization in continuous time.           \\ \hline

[78]             & Pachinko Allocation Model, PAM      & 2006           & Gibbs Sampling                                   & \begin{tabular}[c]{@{}l@{}}LDA\\ A directed acyclic graph method\end{tabular} & Capture arbitrary topic correlations    \\ \hline

[79]           & GWN-LDA                             & 2007           & Gibbs sampling                                   & \begin{tabular}[c]{@{}l@{}}LDA\\ Hierarchical Bayesian algorithm\end{tabular} & Probabilistic community profile Discovery in social network                \\ \hline

[80]         & On-line lda (OLDA)                  & 2008           & Gibbs Sampling                                   & \begin{tabular}[c]{@{}l@{}}LDA\\ Empirical Bayes method\end{tabular}          & Tracking  and Topic Detection           \\ \hline

[36]             & MG-LDA                              & 2008           & Gibbs sampling                                   & LDA     & Sentiment analysis in multi-aspect      \\ \hline

[66]              & Labeled LDA                         & 2009           & Gibbs Sampling                                   & LDA     & Producing a labeled document collection.                                   \\ \hline

[69]              & Relational Topic Models             & 2009           & Expectation-maximization (EM)                    & LDA     & Make predictions between nodes , attributes,  links structure              \\ \hline

[81]              & HDP-LDA                             & 2009           & Gibbs Sampling                                   & LDA     &      \\ \hline

[82]              & DiscLDA                             & 2009           & Gibbs Sampling                                   & LDA     & Classification and dimensionality reduction in documents                   \\ \hline

[17]  & GeoFolk    & 2010                                & Gibbs sampling              & LDA      & Content management and retrieval of  spatial information    \\ \hline

[65]  & JointLDA   & 2010                                & Gibbs Sampling              & \begin{tabular}[c]{@{}l@{}}LDA\\ Bag-of-word model\end{tabular}                & Mining multilingual topics                                  \\ \hline

[35] & Topic-Aspect Model (TAM)                      & 2010                                & Gibbs Sampling              & \begin{tabular}[c]{@{}l@{}}LDA\\ SVM\end{tabular}                              & Scientific topic discovery                                  \\ \hline

[86] & Dependency-Sentiment-LDA                      & 2010                                & Gibbs Sampling              & LDA      & Sentiment classification \\ \hline

[27] & TopicXP    & 2010                                &                             & LDA      & Source code analysis     \\ \hline

[30] & constrained-LDA                               & 2010                                & Gibbs Sampling              & LDA      & Opinion Mining and Grouping Product Features                \\ \hline

[40] & PET - Popular Events Tracking                 & 2010                                & Gibbs Sampling              & LDA      & Event analysis in social network                            \\ \hline

[39][40]                          & EDCoW      & 2010                                &                             & \begin{tabular}[c]{@{}l@{}}LDA\\ Wavelet Transformation\end{tabular}           & Event analysis in Twitter                                   \\ \hline

[85] & Bio-LDA.   & 2011                                & Gibbs Sampling              & LDA      & Extract biological terminology                              \\ \hline

[64] & Twitter-LDA                                   & 2011                                & Gibbs Sampling              & \begin{tabular}[c]{@{}l@{}}LDA\\ Author-topic model\\ \\ PageRank\end{tabular} & Extracting topical keyphrases and analyzing Twitter content \\ \hline

[41] & max-margin latent Dirichlet allocation (MMLDA & 2011                                & variational inference       & \begin{tabular}[c]{@{}l@{}}LDA\\ SVM\end{tabular}                              & Image Classification and Annotation                         \\ \hline

[34] & Sentence-LDA                                  & 2011                                & Gibbs sampling              & LDA      & Aspects and sentiment discovery for web review              \\ \hline

[87] & PLDA+      & 2011                                & Gibbs sampling              & \begin{tabular}[c]{@{}l@{}}LDA\\ Weighted round-robin\end{tabular}             & Reduce inter-computer communication time                    \\ \hline

[72] & Dirichlet class language model (DCLM),        & 2011                                & variational Bayesian EM (VB-EM) algorithm                      & speech recognition and exploitation of language models                         & Dirichlet class language model (DCLM),                      \\ \hline

\end{tabular}
}
\end{table}

\end{landscape}

\subsection{ A brief look at past work: Research between 2012 to 2013}
According to Table 2, some of the popular works published between 2012 and 2013 focused on a variety of topics, such as music retrieve [88], opinion and aspect mining [89], Event analysis [38].\\

ET-LDA, in this work, the authors developed a joint Bayesian model that performs event segmentation and topic modeling in one unified framework. In fact, they proposed an LDA model to obtain event's topics and analysis tweeting behaviors on Twitter that called  Event and Tweets LDA (ET-LDA). They employed Gibbs Sampling method to estimate the topic distribution [38]. The advantage of ET-LDA is that can extract top general topic for entire tweet collection, which means that is very relevant and influenced with the event.\\

 Mr. LDA, the authors introduced a novel model and parallelized LDA algorithm in the MapReduce framework that called Mr. LDA. In contrast other approaches which use Gibbs sampling for LDA, this model uses variational inference [71]. LDA-GA, The authors focused on the issue of Textual Analysis in Software Engineering. They proposed an LDA model based on Genetic Algorithm to determine a near-optimal configuration for LDA (LDA-GA), This approach is considered by three scenarios that include: (a) traceability link recovery, (b) feature location, and (c) labeling. They applied the Euclidean distance for measuring distance between documents and used Fast collapsed Gibbs sampling to approximate the posterior distributions of parameters[63].\\

TopicSpam, they proposed a generative lda based topic modeling methods for detecting fake review. Their model can obtain differences among truthful and deceptive reviews. Also, their approach can provide a clear probabilistic prediction about how  probable a review is truthful or deceptive. For evaluation, the authors used 800 reviews of 20 Chicago hotels and showed that TopicSpam slightly outperforms TopicTDB in accuracy.\\

SSHLDA, the authors proposed a semi supervised hierarchical approach based on topic model which goals for exploring new topics in the data space whenever incorporating the information from viewed labels of hierarchical into the modeling process, called SemiSupervised Hierarchical LDA (SSHLDA). They applied labels hierarchy in as basic hierarchy, called Base Tree (BT); then they used hLDA to make topic hierarchy automatically for each leaf node in BT, called Leaf Topic Hierarchy (LTH). One of the benefits from SSHLDA is that, it can incorporate labeled topics into the generative process of documents. In addition, SSHLDA can automatically explore latent topic in data space, and extend existing hierarchy of showed topics.\\

\subsection{A brief look at past work: Research between 2014 to 2015}
According to Table 2, some of the popular works published between 2014 and 2015 focused on a variety of topics, such as: Hash/tag discovery [45-46], opinion mining and aspect mining [28-29, 31-32], recommendation system [45, 47-48].\\

Biterm Topic Modeling (BTM), Topic modeling over short texts is an increasingly important task due to the prevalence of short texts on the Web. Short texts are popular on today's Web, especially with emergence of social media. Inferring topics from large scale short texts becomes critical.  They proposed a novel topic model for short texts, namely the biterm topic model (BTM). This model can well capture topics within short texts by explicitly modeling word co-occurrence patterns in the whole corpus. BTM obtains underlying topics in a set of text-documents and a distribution of global from per topic in each of them with an analysis of the generation of biterms. Their results showed that btm generates discriminative topic representations as well as rather coherent topics in short texts. The main advantage of BTM prevents the data sparsity issue with learning a global topic distribution [96-97].\\

TOT-MMM, introduced a hashtag recommendation that called TOT-MMM, This approach is a hybrid model that combines a temporal clustering component similar to that of the Topics-over-Time (TOT) Model with the Mixed Membership Model (MMM) that was originally proposed for word-citation co-occurrence. This model can capture the temporal clustering effect in latent topics, thereby improving hashtag modeling and recommendations. They developed a collapsed Gibbs sampling (CGS)  to approximate the posterior modes of the remaining random variables [45]. The posterior distribution of latent topic equaling $k$ for the nth hashtag in tweet d is given by:\\

$P(z_{({{d}_{n}})}^{h}=k|z_{..}^{\left( m \right)},z_{-dn}^{\left( h \right)},w_{..}^{\left( m \right)},w_{..}^{\left( h \right)},t_{..}^{\left( . \right)}~\propto \frac{{{\beta }_{h}}+c{{_{-dn,k}^{w_{dn}^{\left( h \right)}}}^{{}}}}{vh\beta h+c_{-dn,k}^{\left( h \right)_{{}}^{{}}}}~\frac{\alpha +c_{-dn,k}^{\left( {{d}_{b}} \right)}+c_{.,k}^{\left( {{d}_{m}} \right)}}{K\alpha +{{N}_{dm}}+{{N}_{db}}-1}$

$\times ~\frac{t_{d}^{{{\text{ }\!\!\psi\!\!\text{ }}_{k1}}-1}{{\left( 1-{{t}_{d}} \right)}^{{{\text{ }\!\!\psi\!\!\text{ }}_{k2}}-1}}}{B\left( {{\text{ }\!\!\psi\!\!\text{ }}_{k1}},{{\text{ }\!\!\psi\!\!\text{ }}_{k2}} \right)}$

(h) where denotes the number of hashtags type $w_{dn}^{\left( b \right)}$ assigned to latent topic k, excluding the hashtag currently undergoing processing; $c_{-dn,k}^{\left( h \right)_{{}}^{{}}}$ denotes the number of hashtags assigned to latent topic k, excluding the assignment at position ${{d}_{n}}$;  $c_{-dn,k}^{\left( d{{h}_{{}}} \right)}$ denotes the number of hashtags assigned to latent topic k in tweet d, excluding the hashtag currently undergoing processing; $c_{.,k}^{\left( {{d}_{m}} \right)}$denotes the number of words assigned to latent topic k in tweet  $d;V_{b}^{{}}$ is the number of unique hashtags; $t_{d}^{_{{}}}$ is the tweet time stamp omitting position subscripts and superscripts (all words and hashtags share the same time stamp); ${{\text{ }\!\!\psi\!\!\text{ }}_{k1}},{{\text{ }\!\!\psi\!\!\text{ }}_{k2}}$  are the parameters of the beta distribution for latent topic k.

The probability for a hashtag given the observed words and time stamps is:

$p\left( w_{vn}^{h}\text{ }\!\!|\!\!\text{ }w_{v.}^{\left( m \right)},~{{\text{t}}_{\text{v}}} \right)=~\int p\left( w_{vn}^{\left( h \right)}\text{ }\!\!|\!\!\text{ }{{\theta }^{\left( v \right)}} \right)p\left( {{\theta }^{\left( v \right)}}\text{ }\!\!|\!\!\text{ }w_{v.}^{|\left( m \right)},~{{t}_{v}} \right)d{{\theta }^{\left( v \right)}}$

$s\approx \frac{1}{\overset{\acute{\ }}{\mathop{S}}\,}\underset{s=1}{\overset{\overset{\acute{\ }}{\mathop{S}}\,}{\mathop \sum }}\, \underset{k=1}{\overset{K}{\mathop \sum }}\,\varnothing _{h,k,w_{vn}^{\left( h \right)}}^{\left( s \right)}\theta _{k}^{\left( v \right)\left( s \right)},$

 where $S$ is the total number of recorded sweeps, and the superscript s marks the parameters computed based on a specific recorded sweep. To provide the top N predictions, they ranked  from largest to smallest and output the first $N$ hashtags.\\

rLDA , the authors introduced a novel probabilistic formulation to obtain the relevance of a tag with considering all the other images and their tags and also they proposed a novel model called regularized latent Dirichlet allocation (rLDA). This model can estimates the latent topics for each document, with making use of other documents. They used a collective inference scheme to estimate the distribution of latent topics and applied a deep network structure to analyze the benefit of regularized LDA [45-46].\\

\subsection{ A brief look from some impressive past works: Research in 2016}
According to Table 2, some of the popular works published for this year focused on a variety of topics, such as recommendation system [42-44], opinion mining and aspect mining [28-32].\\

A bursty topic on Twitter is one that triggers a surge of relevant tweets within a short period of time, which often reflects important events of mass interest. How to leverage Twitter for early detection of bursty topics has, therefore, become an important research problem with immense practical value. In TopicSketch [59],  proposed a sketch-based topic model together with a set of techniques to achieve real-time bursty topic detection from the perspective of topic modeling, that called in this paper TopicSketch.\\

The bursty topics are often triggered by some events such as some breaking news or a compelling basketball game, which get a lot of attention from people, and "\textbf{force}" people to tweet about them intensely. For example, in physics, this "\textbf{force}" can be expressed by "\textbf{acceleration}", which in our setting describes change of "\textbf{velocity}", i.e., arriving rate of tweets. Bursty topics can get significant acceleration when they are bursting, while the general topics usually get nearly zero acceleration. So the "\textbf{acceleration}" trick can be used to preserve information of bursty topics but filter out the others. Equation (3) shows how we calculate the "\textbf{velocity}" $\hat{v}\left( t \right)$ and "acceleration"$\hat{a}\left( t \right)$ of words.\\

\[{{\hat{v}}_{\Delta T}}=\underset{{{t}_{i}}\le t}{\mathop \sum }\,{{X}_{i}}.\frac{\text{exp}\left( \left( {{t}_{i}}-t \right)/\Delta T \right)}{\Delta T}\]\\

In Equation (1), ${{X}_{i}}$ is the frequency of a word (or a pair of words, or a triple of words) in the i-th tweet, ${{t}_{i}}$ is its timestamp. The exponential part in ${{\hat{v}}_{\Delta T}}\left( t \right)$ works like a soft moving window, which gives the recent terms high weight, but gives low weight to the ones far away, and the smoothing parameter $\Delta T$ is the window size. In fact, the authors proposed a novel data sketch which efficiently maintains at a cost of low-level computational of three quantities: the total number of every tweets, the occurrence of per word and also the occurrence of per word pair. Therefore, low calculation costs is one of the advantages from this approach.

Hashtag-LDA, the authors a personalized hashtag recommendation approach is introduced according to the latent topical information in untagged microblogs. This model can enhance influence of hashtags on latent topics' generation by jointly modeling hashtags and words in microblogs. This approach inferred by Gibbs sampling to  latent topics and  considered a real-world Twitter dataset to  evaluation their approach [44]. CDLDA proposed a conceptual dynamic latent Dirichlet allocation model for tracking and topic detection for conversational communication, particularly for spoken interactions. This model can extract dependencies between topics and speech acts. The CDLDA  applied hypernym information and speech acts for topic detection and tracking in conversations, and it captures contextual information from transitions, incorporated concept features and speech acts [61].

\begin{landscape}
 \begin{table}[]
\centering
\caption{Some impressive articles based on LDA: between 2012-2016}
\label{my-label}
\resizebox{19cm}{!} {
\begin{tabular}{|l|l|l|l|l|l|}
\hline
\multicolumn{1}{|c|}{\textbf{Author-Study}} & \multicolumn{1}{c|}{\textbf{Model}}                          & \multicolumn{1}{c|}{\textbf{Years}} & \multicolumn{1}{c|}{\textbf{Parameter Estimation / Inference}} & \multicolumn{1}{c|}{\textbf{Methods}}                               & \multicolumn{1}{c|}{\textbf{Problem Domain}}    \\ \hline

[90] & locally discriminative topic model (LDTM) & 2012                                & Expectation-maximization (EM)      & LDA                    & Document semantic analysis               \\ \hline

[38] & ET-LDA                                    & 2012                                & Gibbs sampling                     & LDA                    & Event segmentation Twitter               \\ \hline

[88] & infinite latent harmonic allocation (iLHA)                                   & 2012                                & \begin{tabular}[c]{@{}l@{}}Expectation-maximization (EM) algorithm\\ Variational Bayes (VB)\end{tabular} & \begin{tabular}[c]{@{}l@{}}LDA\\ HDP(Hierarchical Dirichlet processes)\end{tabular}          & multipitch analysis and music information retrieval                         \\ \hline

[71] & Mr. LDA                                   & 2012                                & Variational Bayes inference        & \begin{tabular}[c]{@{}l@{}}LDA\\ Newton-Raphson method\\ \\ MapReduce Algorithm\end{tabular} & Exploring document collections from large scale                             \\ \hline

[91] & FB-LDA , RCB-LDA                          & 2012                                & Gibbs Sampling                     & LDA                    & \begin{tabular}[c]{@{}l@{}}Analyze and track public sentiment variations\\ (on twitter)\end{tabular}           \\ \hline

[92] & factorial LDA                             & 2012                                & Gibbs sampling                     & LDA                    & \begin{tabular}[c]{@{}l@{}}Analysis text in a \\ multi-dimensional structure\end{tabular}                      \\ \hline

[93] & SShLDA                                    & 2012                                & Gibbs sampling                     & \begin{tabular}[c]{@{}l@{}}LDA\\ hLDA\end{tabular}        & Topic discovery in data space            \\ \hline

[94] & Utopian                                   & 2013                                & Gibbs sampling                     & LDA                    & Visual text analytics                    \\ \hline

[63] & LDA-GA                                   & 2013                                & Gibbs sampling                     & \begin{tabular}[c]{@{}l@{}}LDA\\ Genetic Algorithm\end{tabular}                              & Software textual retrieval and analysis  \\ \hline

[33] & Multi-aspect Sentiment Analysis for Chinese Online Social Reviews (MSA-COSRs & 2013                                & Gibbs Sampling?                    & LDA                    & \begin{tabular}[c]{@{}l@{}}Sentiment analysis\\ And aspect mining of \\ \\ Chinese social reviews\end{tabular} \\ \hline

 [89] & TopicSpam                                 & 2013                                & Gibbs sampling                     & LDA                    & opinion spam detection                   \\ \hline

[95] & WT-LDA                                    & 2013                                & Gibbs sampling                     & LDA                    & Web Service Clustering                   \\ \hline

[51] & emotion-LDA(ELDA)                          & 2014                                & Gibbs sampling              & LDA                        & Social emotion classification of online news                             \\ \hline

[48] & TWILITE                                    & 2014                                & EM algorithm                & LDA                        & Recommendation system for Twitter     \\ \hline

[98] & Red-LDA                                    & 2014                                & Gibbs-Samplin               & LDA                        & \begin{tabular}[c]{@{}l@{}}Extract information and and data modeling in\\ Patient Record Notes\end{tabular} \\ \hline

[96, 97] & Biterm-Topic-Modeling(BTM)                & 2014                                & Gibbs sampling              & LDA                        & Document clustering for short text    \\ \hline

[47] & Trend Sensitive-Latent Dirichlet Allocation (TS-LDA)                          & 2014                                & Gibbs sampling              & \begin{tabular}[c]{@{}l@{}}LDA\\ Normalized Discounted Cumulative Gain (nDCG)\\ Amazon Mechanical Turk (AMT)2 platform\end{tabular} & Interesting tweets discover for users, recommendation system             \\ \hline

[32] & Fine-grained Labeled LDA (FL-LDA), Unified Fine-grained Labeled LDA (UFL-LDA) & 2014                                & Gibbs sampling              & LDA                        & Aspect extraction and review mining   \\ \hline

[45, 46]                         & Regularized latent Dirichlet allocation (rLDA)                                & 2014                                & Variational Bayes inference & LDA                        & Automatic image tagging or tag recommendation                            \\ \hline

[29] & generative probabilistic aspect mining model (PAMM)                           & 2014                                & Expectation-maximization (EM)                                  & LDA                        & Opinion mining and groupings of drug reviews, aspect mining              \\ \hline

[28] & AEP-based Latent Dirichlet Allocation (AEP-LDA)                               & 2014                                & Gibbs sampling              & LDA                        & Opinion /aspect mining and sentiment word identification                 \\ \hline

[31] & ADM-LDA                                    & 2014                                & Gibbs sampling              & \begin{tabular}[c]{@{}l@{}}LDA\\ Markov chain\end{tabular}    & Aspect mining and sentiment analysis  \\ \hline

[99] & MRF-LDA     & 2015     & EM algorithm & \begin{tabular}[c]{@{}l@{}}LDA\\ Markov Random Field\end{tabular}                                & Exploiting word correlation knowledge                                    \\ \hline
[100] & LightLDA           & 2015 & Gibbs Sampling & LDA    & Topic modeling for very large data sizes\\ \hline
[101]& LFLDA,LF-DMM & 2015 & Gibbs Sampling & LDA & Document clustering for short text \\ \hline   

[103]& LFT          &  2015 & Gibbs Sampling & LDA    & Semantic community detection      \\ \hline             [102] & ATC          & 2015  & EM             & LDA    & Author community discovery    \\ \hline                [106] & FLDA, DFLDA  & 2015  & Gibbs Sampling & LDA    & Multi-label document categorization  \\ \hline
{[}41{]} & Hashtag-LDA               & 2016                                & Gibbs sampling              & LDA                              & Hashtag recommendation and Find relationships between topics and hashtags          \\ \hline

[44] & PMB-LDA                   & 2016                                & Expectation-maximization (EM)                                  & LDA                              & Extract the population mobility behaviors for large scale                          \\ \hline

[108] & Automatic Rule Generation (LARGen                            & 2016                                & Gibbs Sampling              & LDA                              & Malware analysis and Automatic Signature Generation                                \\ \hline

{[}99{]} & PT-LDA                    & 2016                                & Gibbs-EM algorithm          & LDA                              & Personality recognition in social network       \\ \hline

[110]                                  & Corr-wddCRF               & 2016                                & Gibbs sampling              & LDA                              & Knowledge Discovery in Electronic Medical Record                                   \\ \hline

[42]     & multi-idiomatic LDA model (MiLDA)                            & 2016                                & Gibbs sampling              & LDA                              & Content-based recommendation and automatic linking                                 \\ \hline

[43] & Location-aware Topic Model (LTM)                             & 2016                                & Gibbs sampling              & LDA                              & Music Recommendation                            \\ \hline

[62] & TopPRF                    & 2016                                & Gibbs sampling              & LDA                              & Evaluate the relevancy between feedback documents                                  \\ \hline

[50] & contextual sentiment topic model (CSTM)                      & 2016                                & Expectation-maximization (EM)                                  & LDA                              & Emotion classification in social network        \\ \hline

[61] & conceptual dynamic latent Dirichlet allocation (CDLDA)      & 2016                                & Gibbs sampling              & LDA                              & Topic detection in conversations                \\ \hline

[60] & multiple-channel latent Dirichlet allocation (MCLDA)         & 2016                                & Gibbs sampling              & LDA                              & \begin{tabular}[c]{@{}l@{}}Find the relations between diagnoses and medications from  \\ healthcare data\end{tabular} \\ \hline

[37]  & multi-modal event topic model (mmETM)                        & 2016                                & Gibbs sampling              & LDA                              & Tracking  and social event analysis             \\ \hline

[111]                              & Dynamic Online Hierarchical Dirichlet Process model (DOHDP) & 2016                                & Gibbs samplin               & LDA                              & Dynamic topic evolutionary discovery for Chinese social media                      \\ \hline

[59]  & Topicsketch               & 2016                                & Gibbs sampling              & \begin{tabular}[c]{@{}l@{}}LDA\\ Ttensor decomposition algorithm\\ \\ Count-Min algorithm\end{tabular} & Realtime detection and bursty topics dicovery from Twitter                         \\ \hline

[112]                           & fast online EM (FOEM)     & 2016                                & Expectation-maximization (Batch EM)                            & LDA                              & Big topic modeling                              \\ \hline

[113]                                & Joint Multi-grain Topic Sentiment(JMTS)                     & 2016                                & Gibbs sampling              & LDA                              & Extracting semantic aspects  from online reviews                                   \\ \hline

[114]                                   & Character word topic model (CWTM)                            & 2016                                & Gibbs sampling              & LDA                              & Capture the semantic contents in text documents(Chinese language).                 \\ \hline

\end{tabular}
}
\end{table}
\end{landscape}

mmETM, the authors proposed a novel multi modal social event  tracking to capture the evolutionary trends from social events and also for generating effective event summary details over time. The mmETM can model the multimodal property of social event and learn correlations among visual modalities and textual to apart the non-visual-representative topics and visual representative topics. This model can work in an online mode with the event consisting of many stories over time and this is a great advantage for getting the evolutionary trends in event tracking and evolution [37].

\section{Topic Modeling for which the area is used?}
With the passage of time, the importance of Topic modeling in different disciplines will be increase. According to previous studies, we present a taxonomy of current approaches topic models based on LDA model and in different subject such as Social Network [76, 115-118], Software Engineering [8, 9, 25-26], Crime Science [119-121] and also in areas of Geographical [13, 14-17], Political Science [19-20, 122], Medical/Biomedical [4, 123-125] and Linguistic science [126-130] as illustrated by Figure. 2.

\begin{figure*}
 \centering \includegraphics[scale=.7]{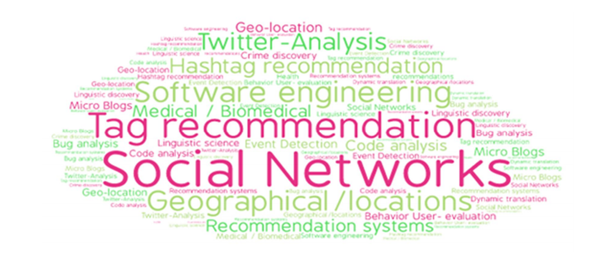}
\caption{ A  vision of the application of Topic modeling in various sciences (based on previous works)}
\label{fig:1}     % Give a unique label
\end{figure*}

\subsection{Topic modeling in Linguistic science}
LDA is an advanced textual analysis technique grounded in computational linguistics research that calculates the statistical correlations among words in a large set of documents to identify and quantify the underlying topics in these documents. In this subsection, we examine some of topic modeling methodology from computational linguistic research, shown some significant research in Table 3. In [130], employed the distributional hypothesis in various direction and it efforts to cancel the requirement of a seed lexicon as an essential prerequisite for use of bilingual vocabulary and introduce various ways to identify the translation of words among languages. In [128] introduced a method that leads the machine translation systems to relevant translations based on topic-specific contexts and used the topic distributions to obtain topic-dependent lexical weighting probabilities. They considered a topic model for training data, and adapt translation model. To evaluate their approach, they performed experiments on Chinese to English machine translation and show the approach can be an effective strategy for dynamically biasing a statistical machine translation towards relevant translations.\\

\begin{table}[]
\centering
\caption{Impressive works LDA-based in Linguistic science}
\label{my-label}
\resizebox{10cm}{!} {

\begin{tabular}{|l|l|l|l|}
\hline
\multicolumn{1}{|c|}{\textbf{Study}} & \multicolumn{1}{c|}{\textbf{Year}} & \multicolumn{1}{c|}{\textbf{Purpose}}                                                                                                & \multicolumn{1}{c|}{\textbf{Dataset}}                                                                            \\ \hline

[130]                           & 2011                               & \begin{tabular}[c]{@{}l@{}}Introduce various ways to\\ identify the translation of\\ words among languages {[}BiLDA{]}.\end{tabular} & \begin{tabular}[c]{@{}l@{}}A Wikipedia dataset (Arabic,\\  Spanish, French, \\ Russian and English)\end{tabular} \\ \hline

[129]                            & 2010                               & \begin{tabular}[c]{@{}l@{}}Obtain term weighting\\  based on LDA\end{tabular}                                                        & A multilingual dataset                                                                                           \\ \hline

[127]                            & 2013                               & \begin{tabular}[c]{@{}l@{}}Present a diversity of\\ new visualization techniques\\ to make concept of topic-solutions\end{tabular}   & \begin{tabular}[c]{@{}l@{}}- Dissertation abstracts\\ (1980 to 2010)\\ - 1 million abstracts\end{tabular}           \\ \hline

[126]                            & 2012                               & \begin{tabular}[c]{@{}l@{}}A topic modeling approach,\\ that it consider geographic\\ information\end{tabular}                       & Foursquare Dataset                                                                                               \\ \hline

[131]                            & 2014                               & \begin{tabular}[c]{@{}l@{}}An approach that is capable\\ to find a document with\\ different language\end{tabular}                   & ALTW2010                                                                                                         \\ \hline

[132]                            & 2013                               & \begin{tabular}[c]{@{}l@{}}A method for linguistic discovery\\ and conceptual metaphors resources\end{tabular}                       & Wikipedia                                                                                                        \\ \hline

\end{tabular}
}
\end{table}

In [127] presented a diversity of new visualization techniques to make the concept of topic-solutions and introduce new forms of supervised LDA, to evaluate they considered a corpus of dissertation abstracts from 1980 to 2010 that belongs to 240 universities in the United States. In [126] developed a standard topic modeling approach, that consider geographic and temporal information and this approach used to Foursquare data and discover the dominant topics in the proximity of a city. Also, the researchers have shown that the abundance of data available in location-based social network (LBSN) enables such models to obtain the topical dynamics in urbanite environments. In [132] have introduced a method for discovery of linguistic and conceptual metaphors resources and built an LDA model on Wikipedia; align its topics to possibly source and aim concepts, they used from both target and source domains to identify sentences as potentially metaphorical. In [131] presented an approach that is capable to find a document with a different language and identify the current language in a document and next step calculate their relative proportions, this approach is based on LDA and used from ALTW2010 as a dataset to evaluation their method.

\subsection{Topic modeling in political science}

Some topic modeling methods have been adopted in the political science literature to analyze political attention. In settings where politicians have limited time-resources to express their views, such as the plenary sessions in parliaments, politicians must decide what topics to address. Analyzing such speeches can thus provide insight into the political priorities of the politician under consideration. Single membership topic models that assume each speech relates to one topic; have successfully been applied to plenary speeches made in the 105th to the 108th U.S. Senate in order to trace political attention of the Senators within this context over time [18]. In addition, in [20] proposed a new two-layer matrix factorization methodology for identifying topics in large political speech corpora over time and identify both niche topics related to events at a particular point in time and broad, long-running topics. This paper has focused on European Parliament speeches, the proposed topic modeling method has a number of potential applications in the study of politics, including the analysis of speeches in other parliaments, political manifestos, and other more traditional forms of political texts. In [19], the purpose of the study is to examine the various effects of dataset selection with consideration of policy orientation classifiers and built three datasets that each data set include of a collection of Twitter users who have a political orientation.In this approach, the output of an LDA has been used as one of many features as a feed to apply SVM classifier and another part of this method used an LLDA that Considered as a stand-alone classifier. Their assessment showed that there are some limitations to building labels for non-political user categories. Shown some significant research based on LDA to political science in Table 4.

\begin{table}[]
\centering
\caption{Impressive works LDA-based in political science}
\label{my-label}
\resizebox{10cm}{!} {

\begin{tabular}{|l|l|l|l|}
\hline
\multicolumn{1}{|c|}{\textbf{Study}} & \multicolumn{1}{c|}{\textbf{Year}} & \multicolumn{1}{c|}{\textbf{Purpose}}                                                                                                              & \multicolumn{1}{c|}{\textbf{Dataset}}                                                                                                                                                                                                                                                                             \\ \hline
[19]                             & 2013                               & \begin{tabular}[c]{@{}l@{}}Evaluate the behavioral effects\\ of different databases from\\ political orientation\\ classifiers\end{tabular}        & \begin{tabular}[c]{@{}l@{}}-Political Figures Dataset\\ -Politically Active Dataset\\ -Politically Modest Dataset\\ -Conover 2011 Dataset (C2D)\end{tabular}                                                                                                                                                      \\ \hline

[21]                             &                                    & \begin{tabular}[c]{@{}l@{}}Introduce a topic model for \\ contrastive opinion modeling\end{tabular}                                                & \begin{tabular}[c]{@{}l@{}}Statement records of \\ U.S. senators\end{tabular}                                                                                                                                                                                                                                     \\ \hline

[133]                            & 2012                               & \begin{tabular}[c]{@{}l@{}}Detection topics that evoke\\ different reactions \\ from communities that\\ lie on the political spectrum\end{tabular} & \begin{tabular}[c]{@{}l@{}}A collection of blog posts\\  from five blogs:\\ 1. Carpetbagger(CB) \\ thecarpetbaggerreport.com\\ 2. Daily Kos(DK) \\ dailykos.com\\ 3. Matthew Yglesias(MY)\\ yglesias.thinkprogress.org\\ 4.Red State(RS)\\ redstate.com\\ 5.Right Wing News(RWN)\\ rightwingnews.com\end{tabular} \\ \hline

[18]                             & 2010                               & \begin{tabular}[c]{@{}l@{}}Discover the hidden relationships\\ between opinion word and topics words\end{tabular}                                  & \begin{tabular}[c]{@{}l@{}}The statement records of\\ senators through the\\ Project Vote Smart \\ (http://www.votesmart.org)\end{tabular}                                                                                                                                                                        \\ \hline

[134]                            & 2014                               & \begin{tabular}[c]{@{}l@{}}Analyze issues related to\\ Korea's presidential election\end{tabular}                                                  & \begin{tabular}[c]{@{}l@{}}Project Vote Smart WebSite\\ (https://votesmart.org/)\end{tabular}                                                                                                                                                                                                                     \\ \hline

[135]                            & 2014                               & \begin{tabular}[c]{@{}l@{}}Examine Political Contention\\ in the U.S. Trucking Industry\end{tabular}                                               & Regulations.gov online portal                                                                                                                                                                                                                                                                                     \\ \hline

[136]                            & 2015                               & \begin{tabular}[c]{@{}l@{}}presented a method for \\ multi-dimensional analysis\\ of political documents\end{tabular}                              & \begin{tabular}[c]{@{}l@{}}Three Germannational elections\\ (2002, 2005 and 2009)\end{tabular}                                                                                                                                                                                                                    \\ \hline
\end{tabular}
}
\end{table}

Fang et al. in [21] suggested a new unsupervised topic model based on LDA for contrastive opinion modeling which purpose to find the opinions from multiple views, according to a given topic and their difference on the topic with qualifying criteria, the model called Cross-Perspective Topic (CPT) model. They performed experiments with both qualitative and quantitative measures on two datasets in the political area that include: first dataset is statement records of U.S. senators that show political stances of senators by these records, also for the second dataset, extracts of world News Medias from three representative media in U.S (New York Times), China (Xinhua News) and India (Hindu). To evaluate their approach with other models used corrIDA and LDA as two baselines. Yano et al. in [137, 138] applied several probabilistic models based on LDA to predict responses from political blog posts. In more detail, they used topic models LinkLDA and CommentLDA to generate blog data(topics, words of post) in their method and with this model can found a relationship between the post, the commentators and their responses. To evaluate their model, gathered comments and blog posts with focusing on American politics from 40 blog sites.\\

In [139] introduced a new application of universal sensing based on using mobile phone sensors and used an LDA topic model to discover pattern and analysis of behaviors of people who changed their political opinions, also evaluated to various political opinions for residents of individual , with consider a measure of dynamic homophily that reveals patterns for external political events. To collect data and apply their approach, they provided a mobile sensing platform to capture social interactions and dependent variables of  American Presidential campaigns of John McCain and President Barack Obama in last three months of 2008. In [133] analyzed reactions of emotions  and suggested a novel model Multi Community Response LDA (MCR-LDA) which in fact is a multi-target and for predicting comment polarity from post content used sLDA and support vector machine classification. To evaluate their approach, provided a dataset of blog posts from five blogs that focus on US politics that was made by [137].\\

In [18], the authors suggested a generative model to auto discover of the latent associations between opinion words and topics that can be useful for extraction of political standpoints and used an LDA model to reduce the size of adjective words,  the authors successfully get that sentences extracted by their model and they shown this model can effectively in different opinions. They were focused on statement records of senators that includes 15, 512 statements from 88 senators from Project Vote Smart WebSite. In [134], examined how social and political issues related to South Korean presidential elections in 2012 on Twitter and used an LDA method to evaluate the relationship between topics extracted from events and tweets. In [136], proposed a method for evaluating and comparing documents, based on an extension of LDA, and used LogicLDA and Labeled LDA approaches for topic modeling in their method. They are considered German National Elections since 1990 as a dataset to apply their method and shown that the use of their method consistently better than a baseline method that simulates manual annotation based on text and keywords evaluation.

\subsection{ Topic modeling in medical and biomedical}
Topic models applied to pure biomedical or medical text mining, researchers have introduced this approach into the fields of medical science. Topic modeling could be advantageously applied to the large datasets of biomedical/medical research, shown some significant research based on LDA to analyze and evaluate content in  medical and biomedical in Table 5. In [125] introduced three LDA-like models and found that this model has higher accuracy than the state-of-the-art alternatives. Authors demonstrated that this approach based on LDA could successfully uncover the probabilistic patterns between Adverse drug reaction (ADR) topics and used ADRS database for evaluating their approach. The aim of the authors to predict ADR from a large number of ADR candidates to obtain a drug target. In [4], focused on the issue of professionalized medical recommendations and introduced a new healthcare recommendation system that called iDoctor, that used Hybrid matrix factorization methods for professionalized doctor recommendation. In fact, They adopted an LDA topic model to extract the topics of doctor features and analyzing document similarity. The dataset this article is college from a crowd sourced website that called Yelp. Their result showed that iDoctor can increase the accuracy of health recommendations and it can has higher prediction in users ratings.\\

\begin{table}[]
\centering
\caption{Impressive works LDA-based in medical and biomedical}
\label{my-label}
\resizebox{10cm}{!} {
\begin{tabular}{|l|l|l|l|}
\hline
\multicolumn{1}{|c|}{\textbf{Study}} & \multicolumn{1}{c|}{\textbf{Year}} & \multicolumn{1}{c|}{\textbf{Purpose/problem domain}}                                                                             & \multicolumn{1}{c|}{\textbf{Dataset}}                                                               \\ \hline

[124]                                    & 2017                               & \begin{tabular}[c]{@{}l@{}}Presented three LDA-based\\ models Adverse Drug \\ Reaction Prediction\end{tabular}                   & ADReCS database                                                                                     \\ \hline

[84]                                     & 2011                               & Extract biological terminology                                                                                                   & \begin{tabular}[c]{@{}l@{}}-MEDLINE and Bio-Terms\\ Extraction\\ -Chem2Bio2Rdf\end{tabular}         \\ \hline

[4]                                      & 2017                               & \begin{tabular}[c]{@{}l@{}}User preference distribution \\ discovery and identity \\ distribution of doctor feature\end{tabular} & \begin{tabular}[c]{@{}l@{}}-Yelp Dataset\\ (Yelp.com)\end{tabular}                                  \\ \hline

[7]                                      & 2012                               & \begin{tabular}[c]{@{}l@{}}-Ranking GENE-DRUG\\ -Detecting relationship \\ between gene and drug\end{tabular}                    & \begin{tabular}[c]{@{}l@{}}National Library of\\ Medicine\end{tabular}                              \\ \hline

[6]                                      & 2011                               & \begin{tabular}[c]{@{}l@{}}Analyzing public health \\ information on Twetter\end{tabular}                                        & \begin{tabular}[c]{@{}l@{}}20 disease articles of\\ twitter data\end{tabular}                       \\ \hline

[115]                                    & 2013                               & \begin{tabular}[c]{@{}l@{}}Analysis of Generated Content\\ by User from social networking\\ sites\end{tabular}                   & \begin{tabular}[c]{@{}l@{}}one million English posted\\ from Facebook's server logs\end{tabular}    \\ \hline

[124]                                    & 2013                               & \begin{tabular}[c]{@{}l@{}}Pattern discovery and extraction\\ for Clinical Processes\end{tabular}                                & \begin{tabular}[c]{@{}l@{}}A data-set from Zhejiang Huzhou\\ Central Hospital of China\end{tabular} \\ \hline

[123]                                    & 2011                               & \begin{tabular}[c]{@{}l@{}}Identifying miRNA-mRNA in\\ functional miRNA regulatory modules\end{tabular}                          & Mouse mammary dataset                                                                               \\ \hline

[140]                                    & 2011                               & Extract common relationship                                                                                                      & T2DM Clinical Dataset                                                                               \\ \hline

[5]                                      & 2012                               & \begin{tabular}[c]{@{}l@{}}Extract the latent topic in\\ Traditional Chinese Medicine\\ document\end{tabular}                    & T2DM Clinical Dataset                                                                               \\ \hline

\end{tabular}
}
\end{table}

In [85], the authors suggested an approach based on LDA that called Bio-LDA that can identify biological terminology to obtain latent topics. The authors have shown that this approach can be applied in different studies such as association search, association predication, and connectivity map generation. And they showed that Bio-LDA can be applied to increase the application of molecular bonding techniques as heat maps. In [7] proposed a topic modeling for rating gene-drug relations by using probabilistic KL distance and LDA that called LDA-PKL and showed that the suggested model achieved better than Mean Average Precision (MAP). They found that the presented method achieved a high Mean Average Precision (MAP) to rating and detecting pharmacogenomics(PGx) relations. To analyze and apply their approach used a dataset from National Library of Medicine. In [6], Presented Ailment Topic Aspect Model (ATAM) to the analysis of more than one and a half million tweets in public health and they were focused on a specific question and specific models; "What public health information can be learned from Twitter?".

In [124] introduced an LDA based method to discover patterns of internal treatment for Clinical processes (CPs), and currently, detect these hidden patterns is one of the most serious elements of clinical process evaluation. Their main approach is to obtain care flow logs and also estimate hidden patterns for the gathered logs based on LDA. Patterns identified can apply for classification and discover clinical activities with the same medical treatment. To experiment the potentials of their approach, used a data-set that collected from Zhejiang Huzhou Central Hospital of China. In [123] introduced a model for the discovery of functional miRNA regulatory modules (FMRMs) that merge heterogeneous datasets and it including expression profiles of both miRNAs and mRNAs, using or even without using exploit the previous goal binding information. This model used a topic model based on Correspondence Latent Dirichlet Allocation (Corr-LDA). As an evaluation dataset, they perform their method to mouse model expression datasets to study the issue of human breast cancer. The authors found that their model is mighty to obtain different biologically meaningful models. In [140], the authors had a study on Chinese medical (CM) diagnosis by topic modeling and introduced a model based on Author-Topic model to detect CM diagnosis from Clinical Information of Diabetes Patients, and called Symptom-Herb-Diagnosis topic (SHDT) model. Evaluation dataset has been collected from 328 diabetes patients. The results indicated that the SHDT model can discover herb prescription topics and typical symptom for a bunch of important medical-related diseases in comorbidity diseases (such as; heart disease and diabetic kidney)[140].\\

\subsection{Topic modeling in geographical and locations}
There is a significant body of research on geographical topic modeling. According to past work, researchers have shown that topic modeling based on location information and textual information can be effective to discover geographical topics and Geographical Topic Analysis. Table 6 shown some significant research based on LDA to topic modeling and content analysis in geographical issues. In [16], this article examines the issue of topic modeling to extract topics from geographic information and GPS-related documents. They suggested a new location text method that is a combination of topic modeling and geographical clustering called LGTA (Latent Geographical Topic Analysis). To test their approaches, they collected a set of data from the website Flickr, according to various topics. In [17] introduced a novel method based on multi-modal Bayesian models to describe social media by merging text features and spatial knowledge that called GeoFolk. As a general outlook, this method can be considered as an extension of Latent Dirichlet Allocation (LDA). They used the available standard CoPhIR dataset that it contains an abundance of over 54 million Flickr. The GeoFolk model has ability to be used in quality-oriented applications and can be merged with some models from Web 2.0 social. In [15] proposed a multiscale LDA model that is a combination of multiscale image representation and probabilistic topic model to obtain effective clustering VHR satellite images.\\

\begin{table}[]
\centering
\caption{Impressive works LDA-based in geographical and locations}
\label{my-label}
\resizebox{10cm}{!} {

\begin{tabular}{|l|l|l|l|}
\hline
\multicolumn{1}{|c|}{\textbf{Study}} & \multicolumn{1}{c|}{\textbf{Year}} & \multicolumn{1}{c|}{\textbf{Purpose}}                                                                                                                                                  & \multicolumn{1}{c|}{\textbf{Dataset}}                                          \\ \hline

[17]                             & 2010                               & \begin{tabular}[c]{@{}l@{}}Discovering multi-faceted\\ summaries of documents\end{tabular}                                                                                             & CoPhIR dataset                                                                 \\ \hline

[16]                             & 2011                               & Content management and retrieval                                                                                                                                                       & Flicker Dataset                                                                \\ \hline

[15]                             & 2013                               & \begin{tabular}[c]{@{}l@{}}Semantic clustering in \\ very high resolution panchromatic\\ satellite images\end{tabular}                                                                 & \begin{tabular}[c]{@{}l@{}}A QUICKBIRD image of\\ a suburban area\end{tabular} \\ \hline

[14]                             & 2010                               & \begin{tabular}[c]{@{}l@{}}Data Discovery, Evaluation of \\ geographically coherent linguistic\\ regions and find the relationship\\ between topic variation and regional\end{tabular} & A Twitter Dataset                                                              \\ \hline

[13]                             & 2008                               & \begin{tabular}[c]{@{}l@{}}Geo-located image categorization\\ and georecognition\end{tabular}                                                                                          & 3013 images Panoramio in France                                                \\ \hline

[141]                            & 2015                               & Cluster discovery in geo-locations                                                                                                                                                     & Reuters-21578                                                                  \\ \hline

[142]                            & 2014                               & \begin{tabular}[c]{@{}l@{}}Discovering newsworthy\\ information From Twitter\end{tabular}                                                                                              & A small Twitter Dataset                                                        \\ \hline
\end{tabular}
}
\end{table}

In[14], the authors introduced a model that includes two sources of lexical variation: geographical area and topic, in another word, this model can discover words with geographical coherence in different linguistic regions, and find a relationship between regional and variety of topics. To test their model, they gathered a dataset from the website Twitter and also we can say that,  also can show from an author's geographic location from raw text [15-17].
In [13] suggested a statistical model for classification of geo-located images based on latent representation. In this model, the content of a geo-located database able be visualized by means of some few selected images for each geo-category. This model can be considered as an extension of probabilistic Latent Semantic Analysis (pLSA). They built a database of the geo-located image which contains 3013 images (Panoramio), that is related to southeastern France.

In addition, in  [143] designed a  browsing system (GeoVisNews) and proposed an extended matrix factorization method based on geographic information for ranking locations, location relevance analysis, and obtain intra-relations between documents and locations, called Ordinal Correlation Consistent Matrix Factorization (OCCMF). For evaluation, the authors used a large multimedia news dataset and showed that their system better than Story Picturing Engine and Yahoo News Map.\\

In this work [141], the authors focused on the issue of identifying textual topics of clusters including spatial objects with descriptions of text. They presented combined methods based on cluster method and topic model to discover textual object clusters from documents with geo-locations. In fact, they used a probabilistic generative model (LDA) and the DBSCAN algorithm to find topics from documents. In this paper, they utilized dataset Reuters-21578 as a dataset for Analysis of their methods. In [142], the authors presented a study on characterizing significant reports from Twitter. The authors introduced a probabilistic model to topic discovery in the geographical topic area and this model can find hidden significant events on Twitter and also considered stochastic variational inference (SVI) to apply gradient ascent on the variable objective with LDA. They collected 2,535 geo-tagged tweets from the Upper Manhattan area of New York. that the KL divergence is a good metric for identifying significant tweet event, but for a large dataset of news articles, the result will be negative.

\subsection{Software engineering and topic modeling}
Software evolution and source code analysis can be effective in solving current and future software engineering problems. Topic modeling has been used in information retrieval and text mining where it has been applied to the problem of briefing large text corpora. Recently, many articles have been published for evaluating / mining software using topic modeling based on LDA, Table 7 shown some significant research based on LDA to source code analysis and an other related issues in software engineering. In [8], for the first time, the authors used LDA, to extract topics in source code and perform to visualization of software similarity, In other words, LDA uses an intuitive approach for calculation of similarity between source files with obtain their respective distributions of each document over topics. They utilized their method on 1,555 software projects from Apache and SourceForge that includes 19 million source lines of code (SLOC). The authors demonstrated this approach, can be effective for project organization, software refactoring. In addition, in [9], introduced a new coupling metric based on Relational Topic Models (RTM) that called Relational Topic based Coupling (RTC), that can identifying latent topics and analyze the relationships between latent topic distributions software data. Also, can say that the RTM is an extension of LDA. The authors used thirteen open source software systems for evaluation this metric and demonstrated that RTC has a useful and valuable impact on the analysis of large software systems.\\

The work in [10] focused on software traceability by topic modeling and proposed a combining approach based on LDA model and automated link capture. They utilized their method to several data sets and demonstrated how topic modeling increase software traceability, and found this approach, able to scale for carried larger numbers from artifacts. In [11] studied about the challenges use of topic models to mine software repositories and detect the evolution of topics in the source code, and suggested the apply of statistical topic models (LDA) for the discovery of textual repositories. Statistical topic models can have different applications in software engineering such as bug prediction, traceability link recovery and software evolution. Chen et al. in [26]  used a generative statistical model(LDA) for analyzing source code and find relationships between software defects and software development. They showed LDA can easily scale to large documents and utilized their approach on three large dataset that includes: Mozilla Firefox, and Mylyn, Eclipse. Linsteadet al. in [25] used and utilized Author-Topic models(AT) to analysis in source codes. AT modeling is an extension of LDA model that evaluation and obtain the relationship of authors to topics and applied their method on Eclipse 3.0 source code including of 700,000 code lines and 2,119 source files with considering of 59 developers. They demonstrated that topic models provided the effective and statistical basis for evaluation of developer similarity.\\

The work in [22] introduced a method based on LDA for automatically categorizing software systems, called LACT. For evaluation of LACT, used 43 open-source software systems in different programming languages and showed LACT can categorization of software systems based on the type of programming language. In addition, in [23, 34],  Proposed an approach topic modeling based on LDA model for the purpose of bug localization. Their idea, applied to the analysis of same bugs in Mozilla and Eclipse and result showed that their LDA-based approach is better than LSI to evaluate and analyze of bugs in these source codes. In [144], introduced a topic-specific approach by considering the combination of description and sensitive data flow information and used an advanced topic model based on LDA with GA, for understanding malicious apps, cluster apps according to their descriptions. They utilized their approach on 3691 benign and 1612 malicious application. The authors found Topic-specific, data flow signatures are very efficient and useful in highlighting the malicious behavior.\\

\begin{table}[]
\centering
\caption{Impressive works LDA-based in software engineering}
\label{my-label}
\resizebox{10cm}{!} {

\begin{tabular}{|l|l|l|l|}
\hline
\multicolumn{1}{|c|}{\textbf{Study}} & \multicolumn{1}{c|}{\textbf{Year}} & \multicolumn{1}{c|}{\textbf{Purpose}}                                                                             & \multicolumn{1}{c|}{\textbf{Dataset}}                                                       \\ \hline

[8]                              & 2007                               & \begin{tabular}[c]{@{}l@{}}Mining software and extracted\\ concepts from code\end{tabular}                        & \begin{tabular}[c]{@{}l@{}}SourceForge and\\ Apache(1,555 projects)\end{tabular}            \\ \hline

[9]                              & 2010                               & \begin{tabular}[c]{@{}l@{}}Identifying latent topics and\\ find their relationships\\ in source code\end{tabular} & \begin{tabular}[c]{@{}l@{}}Thirteen open source\\ software systems\end{tabular}             \\ \hline

[10]                              & 2010                               & Generating traceability links                                                                                     & \begin{tabular}[c]{@{}l@{}}ArchStudio software \\ project\end{tabular}                      \\ \hline

[26]                           & 2012                               & \begin{tabular}[c]{@{}l@{}}Find relationship between the \\ conceptual concerns in source code.\end{tabular}      & Source code entities                                                                        \\ \hline

[25]                             & 2008                               & Analyzing Software Evolution                                                                                      & \begin{tabular}[c]{@{}l@{}}Open source Java projects,\\ Eclipse and ArgoUML\end{tabular}    \\ \hline

[22]                             & 2009                               & \begin{tabular}[c]{@{}l@{}}Automatic Categorization of\\ Software systems\end{tabular}                            & 43 open-source software systems                                                             \\ \hline

[23]                             & 2008                               & \begin{tabular}[c]{@{}l@{}}Source code retrieval for\\ bug localization\end{tabular}                              & Mozila, Eclipse source code                                                                 \\ \hline

[24]                             & 2010                               & \begin{tabular}[c]{@{}l@{}}Automatic bug localization\\ and evaluate its effectiveness\end{tabular}               & \begin{tabular}[c]{@{}l@{}}Open source software such as\\ (Rhino, and Eclipse)\end{tabular} \\ \hline

[144]                            & 2017                               & \begin{tabular}[c]{@{}l@{}}Detection of malicious\\ Android apps\end{tabular}                                     & 1612 malicious application                                                                  \\ \hline
\end{tabular}
}
\end{table}

\subsection{Topic modeling in social network and microblogs}

Social networks are a rich source for knowledge discovery and behavior analysis. For example, Twitter is one of the most popular social networks that its evaluation and analysis can be very effective for analyzing user behavior and etc. Figure 3, shows a simple idea based on LDA algorithm to building a tag recordation system on Twitter. Recently, researchers have proposed many LDA approaches for analyzing user tweets on Twitter. Weng et al. the authors were concentrated on identifying influential twitterers on Twitter and proposed an approach based on an extension of PageRank algorithm to rate users, called TwitterRank, and used an LDA model to find latent topic information from a large collection of documentation. For evaluation this approach, they prepared a dataset from Top-1000 Singapore-based twitterers, showed that their approach is better than other related algorithms [145]. Hong et al.  This paper examines the issue of identifying the Message popularity as measured based on the count of future retweets and sheds. The authors utilized TF-IDF scores and considered it as a baseline, also used Latent Dirichlet Allocation (LDA) to calculate the topic distributions for messages. They collected a dataset that includes 2,541,178 users and 10,612,601 messages and demonstrated that this method can identify messages which will attract thousands of retweets [146]. Table 8 shown some significant research based on LDA to topic modeling and user behavior analysis in social network and microblogs.\\
\begin{figure*}
 \includegraphics[scale=.5]{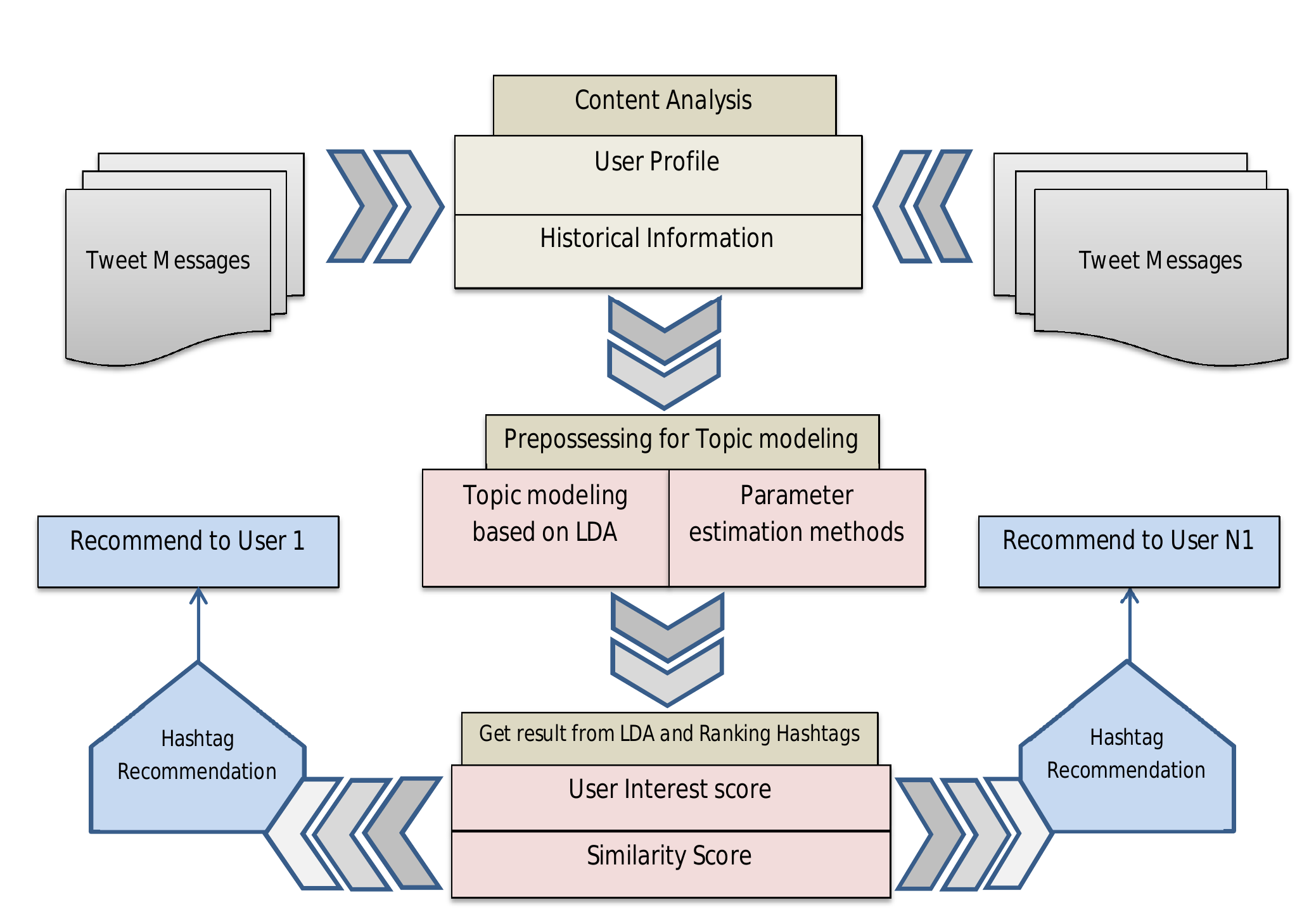}
\caption{A simple framework based on LAD to generate tag as a recommendation system on Twitter }
\label{fig:1}     % Give a unique label
\end{figure*}

The authors in [147] focused on topical recommendations on tweeter and presented a novel methodology for topic discovery of interests of a user on Twitter. In fact, they used a Labeled Latent Dirichlet Allocation (L-LDA) model to discover latent topics between two tweet-sets. The authors found that their method could be better than content based methods for discovery of user-interest. In addition, in [48] suggested, a recommendation system based on LDA for obtaining the behaviors of users on Twitter, called TWILITE. In more detail, TWILTW can calculate the topic distributions of users to tweet messages and also they introduced ranking algorithms in order to recommend top-K followers for users on Twitter.

\begin{table}[]
\centering
\caption{Impressive works LDA-based in social Network and microblogs}
\label{my-label}
\resizebox{10cm}{!} {
\begin{tabular}{|l|l|l|l|}
\hline
\multicolumn{1}{|c|}{\textbf{Study}} & \multicolumn{1}{c|}{\textbf{Year}} & \multicolumn{1}{c|}{\textbf{purpose}}                                                                              & \multicolumn{1}{c|}{\textbf{Dataset}}                                          \\ \hline

[145]                            & 2010                               & \begin{tabular}[c]{@{}l@{}}Finding influential twitterers\\ on social network(Twitter)\end{tabular}                & \begin{tabular}[c]{@{}l@{}}Top-1000 Singapore-based \\ twitterers\end{tabular} \\ \hline

[147]                            & 2014                               & \begin{tabular}[c]{@{}l@{}}Building a topical \\ recommendation systems\end{tabular}                               & A twitter dataset                                                              \\ \hline

[48]                             & 2014                               & \begin{tabular}[c]{@{}l@{}}A recommendation system\\ for Twitter\end{tabular}                                      & A twitter dataset                                                              \\ \hline

[148]                            & 2012                               & \begin{tabular}[c]{@{}l@{}}Analysis and discovered \\ events on Twitter\end{tabular}                               & A twitter dataset                                                              \\ \hline

[91]                             & 2014                               & \begin{tabular}[c]{@{}l@{}}Analyze public sentiment\\ variations regarding a\\ certain tar on Twitter\end{tabular} & A twitter dataset                                                              \\ \hline

[49]                             & 2012                               & \begin{tabular}[c]{@{}l@{}}Analysis of the emotional\\ and stylistic distributions\\ on Twitter\end{tabular}       & A twitter dataset                                                              \\ \hline

[149]                           & 2016                               & \begin{tabular}[c]{@{}l@{}}A topic-enhanced word embedding\\ for Twitter sentiment classification\end{tabular}     & SemEval-2014                                                                   \\ \hline

[150]                            & 2016                               & \begin{tabular}[c]{@{}l@{}}Categorize emotion tendency\\ on Sina Wibo\end{tabular}                                 & A Sina Wibo dataset
\\ \hline

{[}151{]}    & 2016    & \begin{tabular}[c]{@{}l@{}}Ranking items and summarize news\\ information based on hLDA\\  and maximum spanning tree\end{tabular}     & A large news Dataset(CNN,BBC,ABC and Google News)
\\ \hline
\end{tabular}
}
\end{table}

In [121] investigated in the context of a criminal incident prediction on Twitter. They suggested an approach for analysis and understanding of Twitter posts based a probabilistic language model and also considered a generalized linear regression model. Their evaluation showed that this approach is the capability of predict hit-and-run crimes, only using information that exists in the content of the training set of tweets. This paper [152], they introduced a novel method based LDA model to hashtag recommendation on Twitter that can categories posts with them (hashtags). Lin et al. in [153] investigated the cold-start issue with useing the social information for App recommendation on Twitter and used an LDA model to discovering latent group from "Twitter personalities" to recommendations discovery. For test and experiment, they considered Apple's iTunes App Store and Twitter as a dataset. Experimental results show, their approach significantly better than other state-of-the-art recommendation techniques.\\

The authors in [148] presented a technique to analysis and discovered events by an LDA model.The authors found that this method can detect events in inferred topics from tweets by wavelet analysis. For test and evaluation, they collected 13.6 million tweets from Twitter as a dataset and showed the use of both hashtag names and inferred topics is a beneficial effect in description information for events. In addition, in [154] investigated the issue of how to effectively discover and find health-related topics on Twitter and presented an LDA model for identifies latent topic information from a dataset and it includes 2,231,712 messages from 155,508 users. They found that this method may be a valuable tool for detect public health on Twitter. Tan and et al. in [91] focused on tracking public sentiment and modeling on Twitter. They suggest a topic model approach based on LDA, Foreground and Background LDA to distill topics of the foreground. Also proposed another method for ranking a set of reason candidates in natural language, called Reason Candidate and Background LDA (RCB-LDA). Their results showed that their models can be used to identify special topics and find different aspects. The authors in [49] collected a large corpus from Twitter in seven emotions that includes; disgust,   Anger, Fear, Love, Joy, sadness, and surprise. They used a probabilistic topic model, based on LDA, which considered for discovery of emotions in a corpus of Twitter conversations. Srijith et al. in [155] proposed a probabilistic topic model based on hierarchical Dirichlet processes (HDP)) for detection of sub-story. They compared HDP with spectral clustering (SC) and locality sensitive hashing (LSH) and showed that HDP is very effective for story detection data sets, and has an improvement of up to 60\% in the F-score.\\

In [149] proposed a method based on Twitter sentiment classification using topic-enhanced word embedding and also used an LDA model to generate a topic distribution of tweets, considered SVM for classifying tasks in sentiment classification. They used the dataset on SemEval-2014 from Twitter Sentiment Analysis Track. Experiments show that their model can obtain 81.02\% in macro F-measure.In [156]  focused on examining of demographic characteristics in Trump Followers on Twitter. They considered a negative binomial regression model for modeling the "likes" and used LDA to extract tweets of Trump. They provided evaluations on the dataset US2016 (Twitter) that include a number of followers for all the candidates in the United States presidential election of 2016. The authors demonstrated that topic-enhanced word embedding is very impressive for classification of sentiment on Twitter.

\subsection{Crime prediction/evaluation}
Over time; definitely, provides further applications for modeling in various sciences. According to recent work, some researchers have applied the topic modeling methods to crime prediction and analysis. Table 9 shown some significant research based on LDA to topic discovery and crime activity analysis. In [119] introduced an early warning system to find the crime activity intention base on an LDA model and collaborative representation classifier (CRC).The system includes two steps: They utilized LDA for learning features and extract the features that can represent from article sources. For the next step, used from achieved features of LDA to classify a new document by collaborative representation classifier (CRC). In [120] used a statistical topic modeling based on LDA to identify discussion topics among a big city in the United States and used kernel density estimation (KDE) techniques for a standard crime prediction . Sharma et al.  the authors introduced an approach based on the geographical model of crime intensities to detect the safest path between two locations and used a simple Naive Bayes classifier based on features derived from an LDA model [157].

\begin{table}[]
\centering
\caption{Impressive works LDA-based to crime prediction}
\label{my-label}
\resizebox{10cm}{!}{

\begin{tabular}{|l|l|l|l|}
\hline
\multicolumn{1}{|c|}{\textbf{Study}} & \multicolumn{1}{c|}{\textbf{Year}} & \multicolumn{1}{c|}{\textbf{Purpose}}                                                  & \multicolumn{1}{c|}{\textbf{Dataset}}                                                   \\ \hline
[121]                           & 2012                               & \begin{tabular}[c]{@{}l@{}}Automatic semantic analysis\\ on Twitter posts\end{tabular} & \begin{tabular}[c]{@{}l@{}}A corpus of tweets\\ from Twitter(manual)\end{tabular}       \\ \hline

[120]                            & 2014                               & \begin{tabular}[c]{@{}l@{}}Crime prediction using\\ tagged tweets\end{tabular}         & \begin{tabular}[c]{@{}l@{}}City of Chicago Data\\ (data.cityofchicago.org)\end{tabular} \\ \hline
[119]                            & 2015                               & \begin{tabular}[c]{@{}l@{}}Detect the crime activity\\ intention\end{tabular}          & \begin{tabular}[c]{@{}l@{}}800 news articles from\\ yahoo Chinese news\end{tabular}     \\ \hline
\end{tabular}
}
\end{table}

\section{ Open source library and tools / datasets / Software packages and tools for the analysis}
We need new tools to help us organize, search, and understand these vast amounts of information. In this section, we introduce some impressive datasets and tools for Topic Modeling.

\subsection{ Library and Tools}
Many tools for Topic modeling and analysis are available, including professional and amateur software, commercial software, and open source software and also, there are many popular datasets that are considered as a standard source for testing and evaluation. Table 10, show some well-known tools for topic modeling and Table 11, show some well-known datasets for topic modeling. For example; Mallet tools,The MALLET topic model package incorporates an extremely quick  and highly scalable implementation of Gibbs sampling, proficient methods for tools and document-topic hyperparameter optimization for inferring topics for new documents given trained models. Topic models provide a simple approach to analyze huge volumes of unlabeled text. The role of these tools, as mentioned, A "topic" consists of a group of words that habitually happen together. Topic models can associate words with distinguish and similar meanings among uses of words with various meanings and considering contextual clues [158].

\begin{table}[]
\centering
\caption{Some well-known tools for topic modeling}
\label{my-label}
\resizebox{10cm}{!} {
 \begin{tabular}{|l|l|l|l|}
\hline
\multicolumn{1}{|c|}{\textbf{Tools}}                                     & \multicolumn{1}{c|}{\textbf{Implementation/ Language}} & \multicolumn{1}{c|}{\textbf{Inference/Parameter}}                          & \multicolumn{1}{c|}{\textbf{Source code availability}}                                                  \\ \hline

Mallet [159]                                                          & Java                                                   & Gibbs sampling                                                             & http://mallet.cs.umass.edu/topics.php                                                                   \\ \hline

TMT [160]                                                            & Java                                                   & Gibbs sampling                                                             & \begin{tabular}[c]{@{}l@{}}https://nlp.stanford.edu/software/\\ tmt/tmt-0.4/\end{tabular}               \\ \hline

Mr.LDA [71]                                                           & Java                                                   & \begin{tabular}[c]{@{}l@{}}Variational\\ Bayesian\\ Inference\end{tabular} & https://github.com/lintool/Mr.LDA                                                                       \\ \hline

JGibbLDA [161]                                                        & Java                                                   & Gibbs sampling                                                             & http://jgibblda.sourceforge.net/                                                                        \\ \hline

Gensim [162]                                                          & Python                                                 & Gibbs sampling                                                             & https://radimrehurek.com/gensim                                                                         \\ \hline

TopicXP                                                         & Java(Eclipse plugin)                                   &                                                                            & http://www.cs.wm.edu/semeru/TopicXP/                                                                    \\ \hline

\begin{tabular}[c]{@{}l@{}}Matlab Topic\\ Modeling [163] \end{tabular} & Matlab                                                 & Gibbs sampling                                                             & \begin{tabular}[c]{@{}l@{}}http://psiexp.ss.uci.edu/research/\\ programs\_data/toolbox.htm\end{tabular} \\ \hline

Yahoo\_LDA[165]                                                      & C++                                                    & Gibbsampling                                                               & https://github.com/shravanmn/Yahoo\_LDA                                                                 \\ \hline
Lda in R [164]                                                       & R                                                      & Gibbsampling                                                               & \begin{tabular}[c]{@{}l@{}}https://cran.r-project.org/web/\\ packages/lda/\end{tabular}                 \\ \hline
\end{tabular}
}
\end{table}

For evaluation and testing, according to previous work, researchers have released many dataset in various subjects, size, and dimensions for public access and other future work. So, due to the importance of this research, we examined the well-known dataset from previous work. Table 11, shows lists of some famous and popular datasets in various languages.\\

\begin{table}[]
\centering
\caption{Some well-known Dataset for topic modeling}
\label{my-label}
\resizebox{10cm}{!} {

\begin{tabular}{|l|l|l|l|l|}
\hline
\multicolumn{1}{|c|}{\textbf{Dataset}}                                               & \multicolumn{1}{c|}{\textbf{Language}} & \multicolumn{1}{c|}{\textbf{Date of publish}} & \multicolumn{1}{c|}{\textbf{Short-detail}}                                                                                                       & \multicolumn{1}{c|}{\textbf{Availability address}}                                                                          \\ \hline
\begin{tabular}[c]{@{}l@{}}Reuters \\ (Reuters21578)[166]\end{tabular}           & English                                & 1997                                          & \begin{tabular}[c]{@{}l@{}}Newsletters in \\ various categories\end{tabular}                                                                     & \begin{tabular}[c]{@{}l@{}}http://kdd.ics.uci.edu/databases/\\ reuters21578/reuters21578\end{tabular}                       \\ \hline

\begin{tabular}[c]{@{}l@{}}ReutersV 1 \\ (Reuters-Volume I) [167]\end{tabular} & English                                & 2004                                          & \begin{tabular}[c]{@{}l@{}}Newsletters in \\ various categories\end{tabular}                                                                     &                                                                                                                             \\ \hline

UDI-TwitterCrawl-Aug2012 [168]                                                    & English                                & 2012                                          & \begin{tabular}[c]{@{}l@{}}A twitter dataset from\\  millions of tweets\end{tabular}                                                             & \begin{tabular}[c]{@{}l@{}}https://wiki.illinois.edu//wiki/display/\\ forward/Dataset-UDI-TwitterCrawl-Aug2012\end{tabular} \\ \hline

SemEval-2013 Dataset [169]                                                       & English                                & 2013                                          & A twitter dataset from millions of tweets                                                                                                        &                                                                                                                             \\ \hline

Wiki10{[}179{]}                                                                      & English                                & 2009                                          & A Wikipedia Document in various category                                                                                                         & http://nlp.uned.es/social-tagging/wiki10+/                                                                                  \\ \hline

Weibo dataset [170]                                                              & Chinese                                & 2013                                          & A popular Chinese microblogging network                                                                                                          &                                                                                                                             \\ \hline

Bag of Words                                                                & English                                & 2008                                          & \begin{tabular}[c]{@{}l@{}}A multi dataset\\ (PubMed abstracts, \\ KOS blog, \\ NYTimes news, \\ NIPS full papers, \\ Enron Emails)\end{tabular} & \begin{tabular}[c]{@{}l@{}}https://archive.ics.uci.edu/ml/\\ datasets/Bag+of+Words\end{tabular}                             \\ \hline

CiteUlike [171]                                                                  & English                                & 2011                                          & \begin{tabular}[c]{@{}l@{}}A bibliography sharing service\\ of academic papers\end{tabular}                                                      & http://www.citeulike.org/faq/data.adp                                                                                       \\ \hline

DBLP Dataset [172]                                                      & English                                &                                               & \begin{tabular}[c]{@{}l@{}}A bibliographic database about\\ computer science journals\end{tabular}                                               & \begin{tabular}[c]{@{}l@{}}https://hpi.de/naumann/projects/\\ repeatability/\\ datasets/dblp-dataset.html\end{tabular}      \\ \hline

HowNet lexicon                                                                       & Chinese                                & 2000-2013                                     & \begin{tabular}[c]{@{}l@{}}A Chinese machine-readable dictionary\\ / lexical knowledge\end{tabular}                                              & http://www.keenage.com/html/e\_index.html                                                                                   \\ \hline

Virastyar , Persian lexicon [173]                                                 & Persian                                & 2013                                          & Persian poems electronic lexica                                                                                                                  & \begin{tabular}[c]{@{}l@{}}http://ganjoor.net/\\ http://www.virastyar.ir/data/\end{tabular}                                 \\ \hline

NIPS abstracts                                                                       & English                                & 2016                                          & \begin{tabular}[c]{@{}l@{}}The distribution of words \\ in the full text of \\ the NIPS conference (1987 to 2015)\end{tabular}                   & \begin{tabular}[c]{@{}l@{}}https://archive.ics.uci.edu/ml/datasets/\\ NIPS+Conference+Papers+1987-2015\end{tabular}         \\ \hline

\begin{tabular}[c]{@{}l@{}}Ch-wikipedia [114][174]\end{tabular}            & Chinese                                &                                               & A Chinese corpus from Chinese Wikipedia                                                                                                          &                                                                                                                             \\ \hline

\begin{tabular}[c]{@{}l@{}}Pascal VOC 2007 [175][176]\end{tabular}         & English                                & 2007                                          & A dataset of natural images                                                                                                                      & \begin{tabular}[c]{@{}l@{}}http://host.robots.ox.ac.uk/pascal/VOC/\\ voc2007/\end{tabular}                                  \\ \hline

\begin{tabular}[c]{@{}l@{}}AFP\_ARB corpus [177] \end{tabular}                  & Arabic                                 & 2001                                          & \begin{tabular}[c]{@{}l@{}}A collection of newspaper articless\\ in Arabic from Agence France Presse\end{tabular}                                &                                                                                                                             \\ \hline

\begin{tabular}[c]{@{}l@{}}20Newsgroups4 corpus [178] \end{tabular}             & English                                & 2008                                          & \begin{tabular}[c]{@{}l@{}}Newsletters in \\ various categories\end{tabular}                                                                     & http://qwone.com/$\sim$jason/20Newsgroups/                                                                                  \\ \hline

\begin{tabular}[c]{@{}l@{}}New York Times (NYT)\\ dataset [179]\end{tabular}     & English                                & 2008                                          & \begin{tabular}[c]{@{}l@{}}Newsletters in \\ various categories\end{tabular}                                                                     &                                                                                                                             \\ \hline

\end{tabular}
}
\end{table}

\section{Discussion and seven important issues in challenges }
 There are challenges and discussions that can be considered as future work in topic modeling. According to our studies, some issues require further research, which can be very effective and attractive for the future. In this section, we discussed seven important issues and we found that the following issues have not been sufficiently solved. These are the gaps in the reviewed work that would prove to be directions for future work.

\subsection{Topics Modeling in image processing, Image classification and annotation}

Image classification and annotation are important problems in computer vision, but rarely considered together and need some intelligent approach for predict classes.  For example, an image of a class highway is more likely annotated with words "road" and "traffic", "car "  than words "fish " "scuba" and "boat". In [180]  developed a new probabilistic model for jointly modeling the image, its annotations, and its class label.  Their model behaves the class label as a global description of the image and behaves annotation terms as local descriptions of parts of the image. Its underlying probabilistic hypotheses naturally integrate these sources of information.  They derive an approximate inference and obtain algorithms based on variational ways as well as impressive approximations for annotating and classifying new images and extended supervised topic modeling (sLDA) to classification problems.\\

Lienou and et al. in [181] focused on the problem of an image semantic interpretation of large satellite images and used a topic modeling, that each word in a document considering as a segment of image and a document is as an image. For evaluation, they performed experiments on panchromatic QuickBird images.  Philbin and et al. in [182] proposed a geometrically consistent latent topic model to detect  significant objects, called Latent Dirichlet Allocation (gLDA) and then introduced methods for effectiveness of calculations a matching graph, that images are the nodes and the edge strength in visual content. The gLDA method is able to group images of a specific object despite large imaging variations and can also pick out different views of a single object. In [183] introduced a semi-automatic approach to latent information retrieval. According to the hierarchical structure from the images. They considered a combined investigation using LDA model and invariant descriptors of image region for a visual scene modeling. Wick and et al. in [184] They presented an error correction algorithm using topic modeling based on LDA to Optical character recognition (OCR) error correction. This algorithm including two models: a topic model to calculate the word probabilities and an OCR model for obtaining the probability of character errors. In addition, we can combine Topic models with matrix factorization methods to image understanding, tag assignment and semantic discovery from social image datasets [185-186].\\

\subsection{Audio, Music information retrieval and processing}
According to our knowledge, few research works have been done in music information analysis using topic modeling. For example; in [187] proposed a modified version of LDA to process continuous data and audio retrieval. In this model, each audio document includes various latent topics and considered each topic as a Gaussian distribution on the audio feature data. To evaluate the efficiency of their model, used 1214 audio documents in various categories (such as rain, bell, river, laugh, gun, dog and so on). The authors in [188], focused on estimation and estimation of singing characteristics from signals of audio. This paper introduces a topic modeling to the vocal timbre analysis, that each song is considered as a weighted mixture of multiple topics. In this approach, first extracted features of vocal timbre of polyphonic music and then used an LDA model to estimate merging weights of multiple topics. For evaluation, they applied 36 songs that consist of 12 Japanese singers.

\subsection{Drug safety evaluation and Approaches to improving it}
Understanding safety of drug and performance continue to be critical and challenging issues for academia and also it is an important issue in new drug discovery. Topic modeling holds potential for mining the biological documents and given the importance and magnitude of this issue, researchers can consider it as a future work. Bisgin and et al. in [189] introduced an 'in silico' framework to drug repositioning guided through a probabilistic graphical model, that defined a drug as  a 'document' and a phenotype form a drug as a 'word'. They applied their approach on the SIDER database to estimate phenome distribution from drugs and identified 908 drugs from SIDER with new capacity indications and demonstrated that the model can be effective for further investigations. Yu and et al. in [191]  investigated the issue of drug-induced acute liver failure (ALF) with considering the role of topic modeling to drug safety evaluation, they explored the LiverTox database for drugs discovery with a capacity to cause ALF.  Yang and et al.  introduced an automatic approach based on keyphrase extraction to detect expressions of consumer health, according to adverse drug reaction (ADRs) in social media. They used an LDA model as a Feature space modeling to build a topic space on the consumer corpus and consumer health expressions mining.[189-191].\\

\subsection{Analysis of comments of famous personalities, social demographics}
Public social media and micro-blogging services, most notably Twitter, the people have found a venue to hear and be heard by their peers without an intermediary. As a consequence and helped by the public nature of twitter political scientists now potentially have the means to evaluate and understand narratives that organically form, decline among and spread the public in a political campaign. For this field we can refer to some impressive recent works, for example; Wang and et al. they introduced a framework to derive the topic preferences of Donald Trump's followers on Twitter and used LDA to infer the weighted mixture for each Trump tweet  from topics. In addition, we can refer to [156, 192-194].\\

\subsection{ Group discovery and topic modeling}
Graph mining and social network analysis in large graphs is a challenging problem. Group discovery has many applications, such as understanding the social structure of organizations, uncovering criminal organizations, and modeling large scale social networks in the Internet community. LDA Models can be an efficient method for discovering latent group structure in large networks. In [116], the authors proposed a scalable Bayesian alternative based on LDA and graph to group discovery in a big real-world graph. For evaluation, they collected three datasets from PubMed. In [117], the authors introduced a generative approach using a hierarchical Bayes model for group discovery in Social Media Analysis that called Group Latent Anomaly Detection (GLAD) model. This model merged the ideas from both the LDA model and Mixture Membership Stochastic Block (MMSB) model.\\
\subsection{ User Behavior Modeling}
Social media provides valuable resources to analyze user behaviors and capture user preferences. Since the user generated data (such as users activities, user interests) in social media is a challenge[195-196], using topic modeling techniques(such as LDA) can contribute to an important role for the discovery of hidden structures related to user behavior in social media. Although some topic modeling approaches have been proposed in user behavior modeling, there are still many open questions and challenges to be addressed. For example; Giri et al. in [197] introduced a novel way using an unsupervised topic model for hidden interests discovery of users and analyzing browsing behavior of users in a cellular network that can be very effective for mobile advertisements and online recommendation systems. In addition, we can refer to [198-200].\\
\subsection{ Visualizing topic models}
Although different approaches have been investigated to support the visualization of text in large sets of documents such as machine learning, but it is an open challenge in text mining and visualizing data in big data source. Some of the few studies that have been done, such as [201-205]. Chuang and et al. in [205] proposed a topic tool based on a novel visualization technique to the evaluation of textual topical in topic modeling, called Termite. The tool can visualize the collection from the distribution of topic term in LDA with considering a matrix layout. The authors used two measures for understanding a topic model of the Useful terms that including: "saliency" and "distinctiveness". They used the Kullback-Liebler divergence between the topics distribution determined the term for obtain these measures. This tools can increase the interpretations of topical results and make a legible result.\\

\section{ Conclusion}
Topic models have an important role in computer science for text mining. In Topic modeling, a topic is a list of words that occur in statistically significant methods. A text can be an email, a book chapter, a blog posts, a journal article and any kind of unstructured text. Topic models cannot understand the means and concepts of words in text documents for topic modeling. Instead, they suppose that any part of the text is combined by selecting words from probable baskets of words where each basket corresponds to a topic. The tool goes via this process over and over again until it stays on the most probable distribution of words into baskets which call topics. Topic modeling can provide a useful view of a large collection in terms of the collection as a whole, the individual documents, and the relationships between the documents. In this paper, we investigated scholarly articles highly (between 2003 to 2016) related to Topic Modeling based on LDA in various science. Given the importance of research, we believe this paper can be a significant source and good opportunities for text mining with topic modeling based on LDA for researchers and future works.

\section*{Acknowledgements} \label{sec:1}
This article has been awarded by the National Natural Science Foundation of China (61170035, 61272420, 81674099, 61502233), the Fundamental Research Fund for the Central Universities (30916011328, 30918015103), and Nanjing Science and Technology Development Plan Project (201805036).

%\begin{figure}
%% Use the relevant command to insert your figure file.
%% For example, with the graphicx package use
%  \includegraphics{example.eps}
%\end{figure}
%\begin{figure*}
%% Use the relevant command to insert your figure file.
%% For example, with the graphicx package use
%  \includegraphics[width=0.75\textwidth]{example.eps}
%\end{figure*}
%

%\begin{acknowledgements}
%\end{acknowledgements}
 \nocite{*}


\end{document}